# CONVEX GEOMETRY AND STOICHIOMETRY

JER-CHIN (LUKE) CHUANG

Abstract. We demonstrate the benefits of a convex geometric perspective for questions on chemical stoichiometry. We show that the balancing of chemical equations, the use of "mixtures" to explain multiple stoichiometry, and the half-reaction for balancing redox actions all yield nice convex geometric interpretations. We also relate some natural questions on reaction mechanisms with the enumeration of lattice points in polytopes. Lastly, it is known that a given reaction mechanism imposes linear constraints on observed stoichiometries. We consider the inverse question of deducing reaction mechanism consistent with a given set of linear stoichiometric restrictions.

## Contents

A common question encountered in chemistry is the balancing of a chemical equation, and it has been widely discussed in the chemical education literature (see for example the review article by Herndon[12]). It is well-known in the literature (though perhaps not among chemistry students) that the question admits a linear

The author would like to thank the mathematics faculties at Georgia College & State University and at St. Olaf College for the opportunities to present several of these results at departmental colloquia.

algebraic formulation. Here we examine an aspect previously unexplored, namely a convex geometric approach. We show that after specifying the *species* (i.e. the atoms or compounds involved in a reaction) on each side of a chemical equation, *convex polytopes* (a bounded intersection of half-spaces) may be used to provide a visual illustration for questions of existence and uniqueness of balancings, and though generally applicable, it is particularly effective for chemical equations involving only neutral species where no more than three or four elements in total are represented. More generally, we explore the utility of convex geometric representations in chemical stoichiometry, and this forms the principal theme of the current article.

The convex geometric approach complements well the usual linear algebraic method which we summarize in Section 1, with particular attention to issues regarding existence and uniqueness of solutions. That section concludes with a geometric interpretation for the result of the computation. Section 2 discusses the aforementioned geometric approach to the balancing question and its connection with the algebraic approach. The section includes various examples with diagrams illustrating the potential visual appeal and pedagogical value of this approach. In Section 3 we continue to emphasize the utility of the geometric approach by explaining various chemical practices in this geometric framework. In particular we show that the practice of explaining chemical equations arising in contexts of multiple balances as mixtures of equations with unique balances is mathematically justified. Section 4 discusses adaptations of the geometric approach in the presence of charged species. There is an extended discussion of the "half-reaction method" commonly used to balance oxidation-reduction reactions. The geometric approach nicely elucidates the scope and limits of this practice. Section 5 turns attention from individual equations to reaction mechanisms. There are two major considerations in this section: First, we show how natural existence and enumerative questions about mechanisms lead to the much-studied problem of counting lattice points in polytopes. Second, we investigate the moduli of reaction mechanisms consistent with a given overall reaction and observed stoichiometric linear dependencies. The latter complements the "forward" analysis of Missen and Smith [18] in deducing stoichiometric linear dependencies given a reaction mechanism for an overall reaction. Finally, the Appendix (Section 6) includes a geometric approach to some of the observations in Section 2.

Below are some highlights of the results and perspectives of this paper:

- (Section 2) Questions of existence and uniqueness of balancings for a chemical equation may be phrased in terms of the intersections of convex polytopes. This convex geometric approach is visually appealing, perhaps pedagogically beneficial, and intimately connected to the known linear algebraic formulation of the problem.
- (Proposition 3.7) If a chemical equation admits multiple balancings, then any such balance may be realized as a "mixture" of reactions admitting unique balances.
- (Proposition 4.11) If a redox reaction admits a balance, then the "half-reaction method" will generate a balance for the equation. However, in case of multiple balances, not all possible balances may be obtainable via the "half-reaction" method.

- (Remark 5.3) In cases of multiple balances, minimizing the sum of coefficients in the balance may not uniquely determine a balance.
- (Subsection 5.3) It is known that any given mechanism imposes linear dependencies on observed quantities of species. The "inverse" problem of deducing mechanisms consistent with given observed linear dependencies on species yields a collection of mechanisms parametrized by subspaces that can be explicitly identified given knowledge of reaction intermediates. This does not completely solve the inverse problem since convex geometric and order considerations still need to be imposed. The latter is resolved but the former may be computationally expensive.

Computations on convex polytopes were done with Matthias Franz's Maple package CONVEX[10] with the exception of lattice point enumeration which was computed using Verdoolaege's program BARVINOK[24].

## 1. THE ALGEBRAIC APPROACH

The first subsection summarizes the linear algebraic approach, while the second geometrically interprets the result of the computation.

1.1. **The Algebraic Approach in a Nutshell.** We begin with an example illustrating the connection between balancing chemical equations and linear algebra. Further references and examples are provided by Missen and Smith[16], Blakley[4] and Alberty[1] among many others.

1.1. *Example.* Suppose we want to know all possible ways to balance a reaction involving hypothetical neutral species $XY, YZ, XYZ_2$ where $X, Y, Z$ are distinct elements. The algebraic method associates to each species a 3-vector, in this case

$$\mathbf{v}_{XY} = \begin{pmatrix} 1 \\ 1 \\ 0 \end{pmatrix} \qquad \mathbf{v}_{YZ} = \begin{pmatrix} 0 \\ 1 \\ 1 \end{pmatrix} \qquad \mathbf{v}_{XYZ_2} = \begin{pmatrix} 1 \\ 1 \\ 2 \end{pmatrix}$$

where we have tacitly ordered the elements $X, Y, Z$ so that the first component indicates the number of atoms of $X$, the second the number of $Y$ atoms, etc.. Mass-conservation is then reflected by finding *rational* numbers $a_i$ such that

$$a_1 \mathbf{v}_{XY} + a_2 \mathbf{v}_{YZ} + a_3 \mathbf{v}_{XYZ_2} = 0$$

One just multiplies by the least common denominator of the fractions to obtain an integral solution. Thus, if a balancing exists, then a matrix $M$ with the above vectors as its columns has a non-trivial nullspace $\mathrm{NS}(M)$: that is, the equation $M\mathbf{x} = \mathbf{0}$ does not have just the vector $\mathbf{0} = (0, \ldots, 0)^T$ as a solution. For example, we may set

$$M = \begin{pmatrix} 1 & 0 & 1 \\ 1 & 1 & 1 \\ 0 & 1 & 2 \end{pmatrix}$$

(The ordering of the columns does not affect non-triviality of the nullspace.) One computes that $\mathrm{NS}(M) = \mathbf{0}$ so that no balancing exists. Conversely, if $NS(M) \neq \mathbf{0}$, then since $M$ has integer entries, one may find a basis for $NS(M)$ having only rational entries. Hence, a balancing exists iff $NS(M) \neq \mathbf{0}$ and it is furthermore unique (up to multiplicative factor) precisely if the nullspace is one-dimensional.

We formalize the above as follows: Suppose we have $m$ neutral species involving a total of $n$ elements which we place in some order. To each species we associate a $n$-vector where the $k$-th component is the multiplicity of the $k$-th element in that species. Denoting these *species vectors* by $\mathbf{v}_i$, we seek rational numbers $a_i$ such that

$$(1.2) \qquad \sum_{i=1}^{m} a_i \mathbf{v}_i = \mathbf{0}$$

Or equivalent, let $M = (\mathbf{v}_1 \ldots \mathbf{v}_m)$ be a $n \times m$ matrix with the vectors $\mathbf{v}_i$ as columns. Then, a balancing exists iff $\dim(\mathrm{NS}(M)) > 0$ and the balancing is unique (up to multiplicative factor) iff $\dim(\mathrm{NS}(M)) = 1$.

Note that the method does not indicate which species are reactants and which products. Only mass-conservation is being enforced and no directionality is implied, though for a given solution to Equation (1.2) the species are separated into two collections depending on the sign of the coefficients $a_i$. (Those for which $a_i = 0$ are not involved in the balancing.) For convenience, we will refer to these two groups as "reactants" and "products" *without implying any directionality.*

Now, suppose we stipulate that certain species are "reactants" and the remaining "products." In the case where $\dim(\mathrm{NS}(M)) = 1$, the balancing is unique and the species are partitioned into two groups uniquely. However, for $\dim(\mathrm{NS}(M)) > 1$, certain partitions may not be realizable and even if realizable, the balancing may not be unique (even up to multiplicative factor). Mathematically, we seek rational coefficients $a_i$ in Equation (1.2) such that the signs of those associated to "reactants" are opposite of those associated to "products." Computationally, one may proceed as illustrated in Blakley[4] by computing a basis $\{\mathbf{b}_j\}$ for the nullspace. Since now the components of $\mathbf{b}_j$ index the species, we seek linear combinations of $\mathbf{b}_j$ such that components indexing "reactants" have signs opposite those indexing "products."

**1.3. *Example.*** Suppose we have species $XY, XZ, YZ, XYZ, X_5Y_5Z_2$ partitioned into "reactants" $\{X, Y, XYZ\}$ and "products" $\{XZ, YZ, Z_5Y_5Z_2\}$. Ordering the elements $X, Y, Z$ in that order, and defining

$$(1.4) \qquad M = \begin{pmatrix} 1 & 0 & 1 & 1 & 0 & 5 \\ 0 & 1 & 1 & 0 & 1 & 5 \\ 0 & 0 & 1 & 1 & 1 & 2 \end{pmatrix}$$

we compute that $\dim(\mathrm{NS}(M)) = 3$ with the vectors

(1.5)
$$\mathbf{b}_1 = (0, 1, -1, 1, 0, 0)^T \qquad \mathbf{b}_2 = (1, 0, -1, 0, 1, 0)^T \qquad \mathbf{b}_3 = (-3, -3, -2, 0, 0, 1)^T$$

as one possible basis for the nullspace. Note that the components of $\mathbf{b}_i$ index the species $X, Y, XYZ, XZ, YZ, Z_5Y_5Z_2$ *in this order.* Choosing the "reactants" to have non-positive coefficients, we thus seek linear combinations $c_1\mathbf{b}_1 + c_2\mathbf{b}_2 + c_3\mathbf{b}_3$ such that the first three components are non-positive and the remaining non-negative:

$$c_2 - 3c_3 \leq 0 \qquad\qquad\qquad c_1 \geq 0$$
$$c_1 - 3c_3 \leq 0 \qquad\qquad\qquad c_2 \geq 0$$
$$-c_1 - c_2 - 2c_3 \leq 0 \qquad\qquad\qquad c_3 \geq 0$$

We further require that all resulting six components are rational numbers (see following discussion). A geometric approach for this balancing is presented in Example 2.5 below.

We may formalize the preceding as follows: Suppose we partition the $m$ neutral species into $r$ "reactants" and $p$ "products." After possible re-ordering, we may assume vectors $\mathbf{v}_1, \ldots, \mathbf{v}_r$ correspond to the "reactants" and $\mathbf{v}_{r+1}, \ldots, \mathbf{v}_m$ to the "products," where $r+p = m$. Without loss of generality we may choose "reactants" to have non-positive coefficients. Then, each balancing corresponds to a collection of rational numbers $a_i$ such that $\sum_{i=1}^m a_i \mathbf{v}_i = \mathbf{0}$ now with the added stipulation that $a_1, \ldots, a_r$ be non-positive and $a_{r+1}, \ldots, a_m$ non-negative. Let $M = (\mathbf{v}_1 \ldots \mathbf{v}_m)$ and $\{\mathbf{b}_j\}$ be a basis for the nullspace. Since $M$ is a matrix of integers, we may assume that the components of the basis vectors $\mathbf{b}_j$ are all rational. Then, the desired rational $a_i$ correspond precisely to *rational* linear combinations since the $\mathbf{b}_j$ constitute a basis for $\mathrm{NS}(M)$ and hence are linearly independent:

$$(1.6) \qquad \begin{pmatrix} a_1 \\ \vdots \\ a_m \end{pmatrix} = \sum_{j=1}^k c_j \mathbf{b}_j = (\mathbf{b}_1 \ldots \mathbf{b}_k) \begin{pmatrix} c_1 \\ \vdots \\ c_k \end{pmatrix} \qquad a_1, \ldots, a_r \leq 0 \qquad a_{r+1}, \ldots, a_m \geq 0$$

where $k = \dim(\mathrm{NS}(M))$ and the $c_j$ are rational.

## 1.2. A Geometric Interpretation for the Result of the Algebraic Method.

We provide a geometric interpretation for the system of inequalities in Equation (1.6). Each component $a_i$ defines a linear inequality in the variables $c_j$. Geometrically, this represents a half-space in $\mathbb{R}^k$ passing through the origin, where recall $k = \dim(\mathrm{NS}(M))$. Thus, simultaneous solutions to the system of inequalities in Equation (1.6) are geometrically represented by the intersection locus $\mathcal{Q}$ of $m$ half-spaces in $\mathbb{R}^k$. Such an object is called a *polyhedron* in $\mathbb{R}^k$. Since we want rational $c_j$, the possibilities are precisely the rational points (i.e. points where all coordinates are rational numbers) contained in the polyhedron where two such rational points represent the same balancing (up to multiplicative factor) if they are on the same line through the origin. Thus, the "moduli" of balancings with given "reactants" and "products" is realizable as the image of $\mathcal{Q} \subseteq \mathbb{R}^k$ under its projectivization, hence a subset of rational projective space $\mathbb{P}_{\mathbb{Q}}^k$. We will return to a discussion of $\mathcal{Q}$ near the end of the next section. In summary, we have a geometric representation for all possible balancings of a chemical equation when species for both sides are initially specified.

**1.7.** *Example.* Consider the oxidation of nitric oxide ($NO$) to nitrogen dioxide ($NO_2$). The reactants are $NO, O_3$ and the products are $NO_2, O_2$. Note that this example involves two allotropes of oxygen. We order the elements by $O, N$ in that order. Defining

$$(1.8) \qquad M = \begin{pmatrix} 1 & 3 & 2 & 2 \\ 1 & 0 & 1 & 0 \end{pmatrix}$$

we compute that $\dim(\mathrm{NS}(M)) = 2$ with the vectors

$$(1.9) \qquad \mathbf{b}_1 = (0, -2, 0, 3)^T \qquad \mathbf{b}_2 = (-3, -1, 3, 0)^T$$

as one possible basis for the nullspace. Note that the components of $\mathbf{b}_i$ index the species $NO, O_3, NO_2, O_2$ *in this order*. Choosing the reactants to have non-positive coefficients, we thus seek linear combinations $c_1\mathbf{b}_1 + c_2\mathbf{b}_2$ such that the first two components are non-positive and the remaining non-nonegative:

$$-3c_2 \leq 0 \qquad\qquad 3c_2 \geq 0$$
$$-2c_1 - c_2 \leq 0 \qquad\qquad 3c_1 \geq 0$$

The resulting polyhedron $\mathcal{Q}$ is the set $\{(c_1, c_2) | c_i \geq 0\}$ and each rational point within $\mathcal{Q}$ indexes a balancing. A geometric approach for this balancing is presented in Example 2.7 below.

## 2. Geometric Approach to Balancing

Now we examine the balancing question anew but guided by geometric considerations. Suppose as before we partition $m$ neutral species into $r$ "reactants" and $p$ "products" and that these are represented by the collections $\mathbf{v}_1, \ldots, \mathbf{v}_r$ and $\mathbf{v}_{r+1}, \ldots, \mathbf{v}_m$ respectively, where $r + p = m$. Then, balancings are precisely equalities of a non-negative rational linear combination of the former set of vectors with a non-negative rational linear combination of the latter set:

$$(2.1) \qquad\qquad \sum_{i=1}^{r} r_i \mathbf{v}_i = \sum_{j=r+1}^{m} p_j \mathbf{v}_j$$

where the coefficients $r_i, p_j \geq 0$ are rational.

Recall that the set of all non-negative scalings of a vector is geometrically represented by a ray based at the origin in the direction of the vector. Non-negative linear combinations of a set of vectors determine a mathematical object called the *(polyhedral) cone* spanned by the vectors with the origin as the vertex of the cone. Since we only care about the rational numbers $r_i, p_j$ up to multiplicative factor, we slice both cones simultaneously by a hyperplane intersecting all the *positive $x_i$-axes* for $i = 1, \ldots, n$. This hyperplane then intersects each cone transversely and the intersection loci are bounded polyhedra, i.e. *(convex) polytopes*, which we may call the "*reactant*" and "*product*" *polytopes*. Indeed, each is the convex hull of the points given by the intersection of the various defining rays (either those associated with "reactants" $\{\mathbf{v}_1, \ldots, \mathbf{v}_r\}$ or those with "products" $\{\mathbf{v}_{r+1}, \ldots, \mathbf{v}_m\}$) with the slicing hyperplane. Without loss of generality, we may assume the hyperplane is described by an equation of the form $\mathbf{n} \cdot \mathbf{x} = h$ where $\mathbf{n}, h$ consist entirely of rational numbers. Then, given a choice of "reactants" and "products," a necessary condition for the existence of a balancing is that the intersection of these two polytopes is non-empty. The intersection is another polytope, which we will call the *intersection polytope*, and if non-empty, it necessarily contains a rational point. Conversely, one can show that each rational point in the intersection polytope corresponds to some balancing of a chemical equation with the specified grouping of species. (See the discussion at the end of this section or for a geometric argument see the appendix.)

Since we assume that the coefficients $r_i, p_j \geq 0$ are non-negative, all the components of the scaled vectors $r_i\mathbf{v}_i, p_j\mathbf{v}_j$ are non-negative, and a canonical choice for the hyperplane is $\sum_{l=1}^{n} x_l = 1$. Notice that this hyperplane intersects each positive $x_i$-axis for $i = 1, \ldots, n$ as required. The intersection of this canonical hyperplane with the region $\{(x_1, \ldots, x_n) | x_i \geq 0\}$ is a $(n-1)$-dimensional "triangle," called a $(n-1)$-*simplex*. The coordinates $(x_1, \ldots, x_n)$ sum to unity and are called *barycentric coordinates*. Each species vector $\mathbf{v}_i$ corresponds to a point with barycentric coordinates specifying the proportions of each element within the species. For example, if $X, Y, Z$ are the only elements involved and ordered thus, then the barycentric coordinates for the species $X_2Y_3Z_4$ is $(2/9, 1/3, 4/9)$. Taking a non-chemical example, the RGB-system of colors describes colors in terms of proportions $(r, g, b)$ of red, green, and blue, respectively, where $r + g + b = 1$.

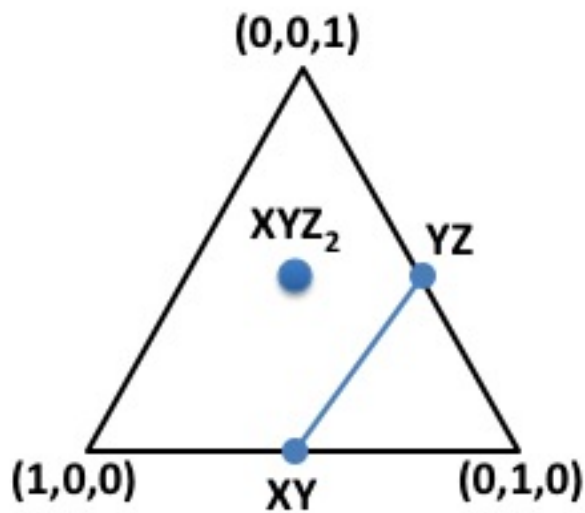

FIGURE 1. Case where no balancing exists for any grouping of reactants and products

Note that rational points on the boundary of the intersection polytope correspond to balancings in which not every species is present. Hence, if we require that all species be present, then if an equation can be balanced then it can be balanced in infinitely many distinct ways unless the intersection polytope is a single point. However, this is only a necessary and not a sufficient condition for the uniqueness of balance. We will see later that additionally we need the generating vectors for the reactant and product cones each to be linearly independent (see Equation 2.18). For example, this is satisfied if the reactant and product polytopes are line segments meeting at a point in the relative interior of each.

We now illustrate this perspective with several examples:

2.2. *Example* (Non-Existence of Balancing). Suppose we have "reactants" $\{XY, YZ\}$ and "product" $\{XYZ_2\}$ where $X, Y, Z$ are distinct elements and all species are neutral. If we order the elements as $X, Y, Z$, then the barycentric coordinates of the "reactant" species are $(1/2, 1/2, 0)^T$ and $(0, 1/2, 1/2)^T$, and that of the "product" species is $(1/4, 1/4, 1/2)^T$. The "reactant" polytope is the convex hull of the barycentric coordinates for the "reactant species," namely a line segment joining $(1/2, 1/2, 0)^T$ and $(0, 1/2, 1/2)^T$ whereas the "product polytope" is just a single point. See Figure 1. Since the line segment and point are disjoint, we conclude that no balancing exists for that particular choice of "reactants" and "products." In fact, from the diagram we can easily see that there is no way to partition the species into "reactants" and "products" to obtain a situation where a balancing exists. This concurs with the conclusion in Example 1.1 obtained via the algebraic method.

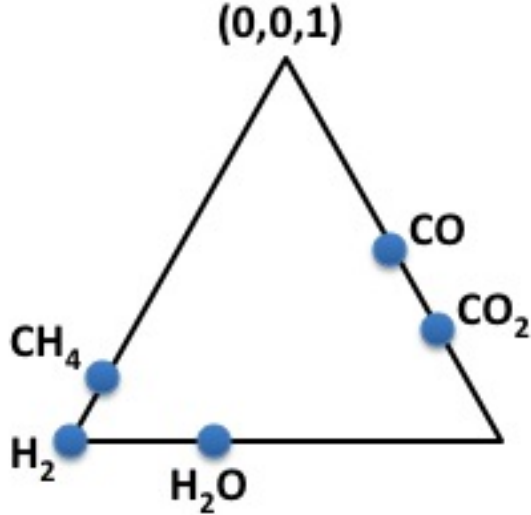

FIGURE 2. Case where uniqueness of balancing depends on choice of reactants and products

2.3. *Example* (Unique and Non-Unique Balancings). Consider the neutral chemical species $H_2, H_2O, CH_4, CO_2, CO$. Ordering the elements $H, O, C$ in that order, the barycentric coordinates are respectively,

$$(2.4) \qquad \begin{pmatrix} 1 \\ 0 \\ 0 \end{pmatrix}, \quad \begin{pmatrix} 2/3 \\ 1/3 \\ 0 \end{pmatrix}, \quad \begin{pmatrix} 4/5 \\ 0 \\ 1/5 \end{pmatrix}, \quad \begin{pmatrix} 0 \\ 2/3 \\ 1/3 \end{pmatrix}, \quad \begin{pmatrix} 0 \\ 1/2 \\ 1/2 \end{pmatrix}$$

See Figure 2. As evident from the diagram, to obtain a non-empty intersection polytope, we need to have at least two "reactants" and two "products" each. By the comments above, we have an unique balancing precisely when four of the five species are involved and are grouped pairwise, and the lines so determined intersect uniquely. One easily checks from the diagram that this yields the groupings:

| | |
|---|---|
| $\{CH_4, CO_2\}$ | $\{H_2, CO\}$ |
| $\{H_2, CO\}$ | $\{CH_4, CO_2\}$ |
| $\{H_2, CO\}$ | $\{CH_4, H_2O\}$ |
| $\{H_2O, CO\}$ | $\{CH_4, CO_2\}$ |
| $\{H_2O, CO\}$ | $\{H_2, CO_2\}$ |

(Compare with cases $(\alpha)$-$(\epsilon)$ in Example 6 of Blakley[4]).

2.5. *Example* (Non-Unique Balancing). Consider "reactants" $\{X, Y, XYZ\}$ and "products" $\{XZ, YZ, X_5Y_5Z_2\}$ where $X, Y, Z$ are distinct elements ordered thus as in Example 1.3. The barycentric coordinates are

$$(2.6) \qquad \begin{pmatrix} 1 \\ 0 \\ 0 \end{pmatrix}, \quad \begin{pmatrix} 0 \\ 1 \\ 0 \end{pmatrix}, \quad \begin{pmatrix} 1/3 \\ 1/3 \\ 1/3 \end{pmatrix}, \quad \begin{pmatrix} 1/2 \\ 0 \\ 1/2 \end{pmatrix}, \quad \begin{pmatrix} 0 \\ 1/2 \\ 1/2 \end{pmatrix}, \quad \begin{pmatrix} 5/12 \\ 5/12 \\ 1/6 \end{pmatrix}$$

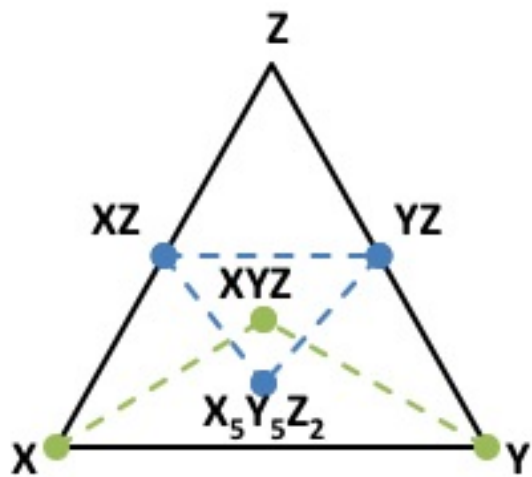

FIGURE 3. Case where infinitely-many distinct balancings exist for given choice of reactants and products

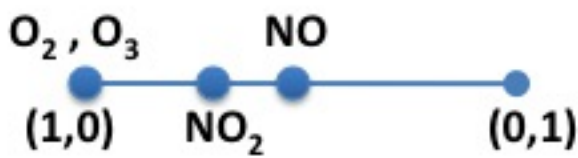

FIGURE 4. Case where allotropes are involved

respectively. See Figure 3. The intersection polytope is a quadrilateral, and hence there are infinitely many distinct balancings under this grouping of species. For example, the rational points $(3/8, 3/8, 1/4)^T$ and $(2/5, 2/5, 1/5)^T$ are in the interior of the intersection polytope with

$$\frac{1}{8}\begin{pmatrix}1\\0\\0\end{pmatrix} + \frac{1}{8}\begin{pmatrix}0\\1\\0\end{pmatrix} + \frac{1}{4}\begin{pmatrix}1\\1\\1\end{pmatrix} = \begin{pmatrix}3/8\\3/8\\1/4\end{pmatrix} = \frac{1}{16}\begin{pmatrix}1\\0\\1\end{pmatrix} + \frac{1}{16}\begin{pmatrix}0\\1\\1\end{pmatrix} + \frac{1}{16}\begin{pmatrix}5\\5\\2\end{pmatrix}$$

$$\frac{1}{5}\begin{pmatrix}1\\0\\0\end{pmatrix} + \frac{1}{5}\begin{pmatrix}0\\1\\0\end{pmatrix} + \frac{1}{5}\begin{pmatrix}1\\1\\1\end{pmatrix} = \begin{pmatrix}2/5\\2/5\\1/5\end{pmatrix} = \frac{1}{40}\begin{pmatrix}1\\0\\1\end{pmatrix} + \frac{1}{40}\begin{pmatrix}0\\1\\1\end{pmatrix} + \frac{3}{40}\begin{pmatrix}5\\5\\2\end{pmatrix}$$

yielding *distinct* balancings:

$$2X + 2Y + 4XYZ = XZ + YZ + X_5Y_5Z_2$$

$$8X + 8Y + 8XYZ = XZ + YZ + 3X_5Y_5Z_2$$

respectively.

2.7. *Example* (Allotropes). Consider the oxidation of nitric oxide ($NO$) to nitrogen dioxide ($NO_2$) as discussed earlier in Example 1.7. The reactants are $NO, O_3$ and the products are $NO_2, O_2$. Ordering the elements by $O, N$ in that order, both allotropes are represented by the barycentric coordinates $(1,0)^T$ whereas those for $NO, NO_2$ are $(1/2, 1/2)^T$ and $(2/3, 1/3)^T$ respectively. The intersection polytope is a line segment $\{(1-t, t)^T | 0 \le t \le 1/3\}$. See Figure 4. Writing $(1-t, t)^T$ for a point in the intersection polytope, we have the representations:

$$(2.8) \qquad \begin{pmatrix} 1-t \\ t \end{pmatrix} = 2t \begin{pmatrix} 1/2 \\ 1/2 \end{pmatrix} + (1-2t) \begin{pmatrix} 1 \\ 0 \end{pmatrix} = t \begin{pmatrix} 1 \\ 1 \end{pmatrix} + \frac{1-2t}{3} \begin{pmatrix} 3 \\ 0 \end{pmatrix}$$

$$(2.9) \qquad \begin{pmatrix} 1-t \\ t \end{pmatrix} = 3t \begin{pmatrix} 2/3 \\ 1/3 \end{pmatrix} + (1-3t) \begin{pmatrix} 1 \\ 0 \end{pmatrix} = t \begin{pmatrix} 2 \\ 1 \end{pmatrix} + \frac{1-3t}{2} \begin{pmatrix} 2 \\ 0 \end{pmatrix}$$

$$(2.10)$$

relative the reactant and product cones, respectively. Thus, we have a one-parameter family of distinct balancings:

$$(2.11) \qquad tNO + \frac{1-2t}{3} O_3 = tNO_2 + \frac{1-3t}{2} O_2$$

indexed by *rational* $t$ such that $0 \le t \le 1/3$. If we insist that all reactants and products be present, then we have strict inequalities. For example, setting $t = 1/4$ and clearing denominators yields the balancing: $6NO + 4O_3 = 6NO_2 + 3O_2$. The case where $t = 1/5$ corresponds to the usual equation for the oxidation: $NO + O_3 \to NO_2 + O_2$.

2.12. *Remark.* In practice, one knows the relative amounts of reactants and products. Exact knowledge of either determines an unique ray in the corresponding cone. Hence, if a solution exists, this information is sufficient to identify an unique point in the intersection polytope. Because of measurement error and uncertainty, we have instead a neighborhood of a ray (mathematically, a neighborhood of a point in *projective space*), and hence possibly infinitely-many balancings. In practice, chemists choose balancings with "small" coefficients, though a mathematical formulation for this "rule-of-thumb" is unclear. See also Remark 5.3..

2.13. *Remark.* Note that we can scale the intersection polytope so that it has integer vertices. Let $d$ be its dimension and write $\nu(n, d)$ for the number of points in the intersection polytope with denominator (in lowest terms) no greater than $n$. Analogously, let $\nu_0(n, d)$ denote the number of such in the interior of the polytope. By convention, we set $\nu(0, d) = 1 = \nu_0(0, d)$. It is known that both $\nu, \nu_0$ are polynomials of degree $d$ in $n$ (see Stanley [21]). In particular, knowledge of $d$ other values determine either $\nu$ or $\nu_0$ completely. For example, we can scale the quadrilateral that is the intersection polytope in Figure 3 to have integer vertices

$$(15, 15, 6)^T \qquad (16, 10, 10)^T \qquad (10, 16, 10)^T \qquad (12, 12, 12)^T$$

Then, with regard to *this* polytope, we can count that in its interior there are 16 integer points and 33 points with denominator at most two so that $\nu_0(n, 2) = n^2 + 14n + 1$. However, relating denominators to the least integral coefficients of a balancing is not direct, as Example 2.7 shows. We will return to enumerative questions in Subsection 5.2.

Now, we describe the connection between the geometric and algebraic approaches: Recall that the algebraic approach parametrizes possible balancings by rational points of an unbounded polyhedron $\mathcal{Q}$ in $\mathbb{R}^k$. Consider the linear transformations

$$(2.14) \qquad B\colon \mathcal{Q} \xrightarrow{(\mathbf{b}_1...\mathbf{b}_k)} \mathbb{R}^m$$

$$(2.15) \qquad C_r\colon \mathbb{R}_+^r \xrightarrow{(\mathbf{v}_1...\mathbf{v}_r)} \mathbb{R}^n$$

$$(2.16) \qquad C_p\colon \mathbb{R}_+^p \xrightarrow{(\mathbf{v}_{r+1}...\mathbf{v}_m)} \mathbb{R}^n$$

where $\mathbb{R}_+^r$ denote non-negative values and we have interpreted matrices as linear maps. Define $R = C_r \circ \pi_r$ and $P = C_p \circ (-\pi_p)$ where $\pi_r$ (resp. $\pi_p$) denotes projection onto the first $r$ (resp. last $p$) coordinates in $\mathbb{R}^m$. Then, $B\colon \mathcal{Q} \to \mathbb{R}^m$ is the centralizer of the maps $R, P\colon \mathbb{R}^m \to \mathbb{R}^n$ (in particular, they coincide on the image of $B$), and $RB, PB\colon \mathcal{Q} \to \mathbb{R}^n$ each map the polyhedron $\mathcal{Q}$ in a linear fashion onto the intersection cone. These maps fit into a commutative diagram:

$$(2.17)$$

where $\tilde{R} = \pi_r \circ B$ and $\tilde{P} = -\pi_p \circ B$.

Let $\mathcal{R}, \mathcal{P}$ be the reactant and product cones, respectively and $\mathcal{I} = \mathcal{R} \cap \mathcal{P}$ the intersection *cone*. Note that $\mathcal{R} = \operatorname{im} C_r$ and $\mathcal{P} = \operatorname{im} C_p$. Define the matrix $M = (\mathbf{v}_1 ... \mathbf{v}_m)$ whose columns are those of $C_r, C_p$ respectively. Note that $\operatorname{NS}(M) = \operatorname{im}(B)$ so that $\dim(NS(M)) = \dim(\mathcal{Q})$ the dimension of the unbounded polyhedron in the algebraic approach. By a fact from linear algebra (see Section 6), one has

$$(2.18) \qquad \dim(\mathcal{Q}) = \dim(\operatorname{NS}(M)) = \dim(\mathcal{I}) + \dim(\ker(C_r)) + \dim(\ker(C_p))$$

so that of the polyhedron $\mathcal{Q}$ from the algebraic approach is not in general linearly isomorphic to the intersection cone. It is so iff $C_r, C_p$ are injective, that is, the reactant and product species vectors are each linearly independent sets. One example where this fails is a reaction involving either more reactant or product species than elements involved.

Intuitively, the nullspace $\operatorname{NS}(M)$ measures linear relations among the reactant and product species vectors. These may be among reactant vectors only, among products vectors only, or among vectors drawn from both reactant and product species, hence the three terms in Equation 2.18. Letting $s, r, p$ be the number of species, reactants, and products respectively, we have in species space $\mathcal{S} = \mathbb{R}^s$ the $(s-1)$-simplex $\Delta^{s-1} = \Delta^{r-1} * \Delta^{p-1}$ where $\Delta^{r-1}, \Delta^{p-1}$ are mutual cofaces defined by the reactant and product vectors, respectively. The species vectors define partial maps $\Delta^{r-1}, \Delta^{p-1} \rightrightarrows \mathcal{E}$ whose images are the reactant and product polytopes, respectively. The bijectively image of the moduli polyhedron $\mathcal{Q}$ in species space $\mathcal{S}$ is possibly "partially collapsed" under the "elemental proportion" map $M$ into the intersection cone in element space $\mathcal{E}$ due to possible linear dependencies among generating rays for the reactant cone and also for the product cone. In summary, under the geometric approach, we gain often a visual determination of existence issues at the expense of possibly losing count of the algebraic dimension of possible multiple balancings and hence a slightly more involved criterion for uniqueness.

## 3. POLYTOPAL EXPLANATION FOR SOME CHEMICAL PRACTICES

In this section we examine several common practices in chemical stoichiometry and show that they all have simple polytopal explanations.

### 3.1. Inspection Methods.

In the chemical education literature, there are numerous "inspection methods" for balancing chemical equations (for example [23],[15]). Mathematically, these often amount to establishing an ad hoc elimination order on the elements, or in geometric terms, matching the projections of the reactant and product polytopes onto various axial directions. For example, the first two of "Ling's Rules for Balancing Redox Equations by Inspection" as presented in Kolb[15] are:

- "Step 1: Locate any elements that must have the same coefficient in the balanced equation, those appearing only *once* on each side of the equation and in *equal numbers* on both sides. Mark these terms with arrows."

- "Step 2: Locate any elements that appear only *once* on each side of the equation but have *unequal* numbers of atoms. Balance these elements first."

This is illustrated in the following example:

3.1. *Example.* [15] In balancing

$$S + HNO_3 \rightarrow SO_2 + NO + H_2O$$

by step 1, we see that $S, SO_2$ have identical coefficents and so also $HNO_3, NO$. By step 2, we balance hydrogen, which forces the balance for nitrogen, then sulfur, then oxygen yielding:

$$\frac{3}{2}S + 2HNO_3 \rightarrow \frac{3}{2}SO_2 + 2NO + H_2O$$

Kolb's article[15] even presents a case where Ling's method does not work, and she resorts to linear algebra, the proper algebraic framework.

### 3.2. Stoichiometric Restriction.

Sometimes one knows from experiment that certain species react in particular ratios, usually from kinetic and mechanistic considerations (see Section 5 for details). Missen and Smith[17] provide an example involving a permanganate-peroxide reaction in acidified aqueous solution:

$$H_2SO_4 + KMnSO_4 + H_2O_2 \rightarrow O_2 + H_2O + K_2SO_4 + MnSO_4 \quad \text{(unbalanced)}$$

One checks (for example via CONVEX[10] or POLYMAKE[11]) that the intersection polytope is 1-dimensional. However, it is known from experiment that $KMnO_4$ reacts with $H_2O_2$ in a $2:5$ ratio. Algebraically, this may be reflected by augmenting an additional *row* to the matrix $M$ mapping species to elements. For example, ordering the species as they appear left-right in the above equation, the augmented matrix is

$$\begin{pmatrix} & & & M & & & \\ 0 & -2 & 5 & 0 & 0 & 0 & 0 \end{pmatrix}$$

Geometrically, we incorporate the restriction by replacing the generating vertices $\mathbf{v}_{KMnSO_4}, \mathbf{v}_{H_2O_2}$ with a linear combination $(2/7)\mathbf{v}_{KMnSO_4} + (5/7)\mathbf{v}_{H_2O_2}$. One can compute that this alteration reduces the dimension of the reactant polytope. Either way, we compute that the intersection polytope is now 0-dimensional so that there is an unique balance under this stoichiometric restriction. Similarly, if $\mathbf{v}_i$ is the generating vertex indexed by the $i$-th specie, and it is known experimentally that reactants $R_{i_1}, \ldots, R_{i_k}$ react in a ratio of $r_{i_1} : \cdots : r_{i_k}$, then we replace the associated generating vertices with the single vertex given as a weighted linear combination:

$$(3.2) \qquad \mathbf{v} = \sum_{j=1}^{k} \frac{r_{i_j}}{r} \mathbf{v}_{i_j} \qquad r = \sum_{j=1}^{k} r_{i_j}$$

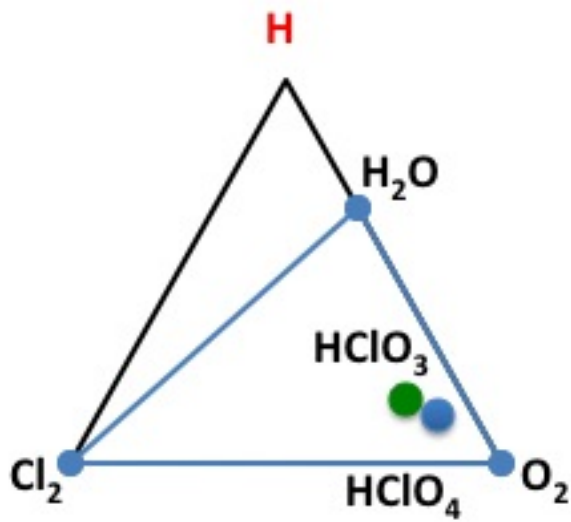

Figure 5. Reaction as Mixture of Reactions

A similar observation applies if certain products are know to be produced in particular ratios. The intersection from the resulting polytopes may or may not be of lower dimension.

More generally, if there is linear relation among the species, it defines a hyperplane $\mathcal{H}$ that interesects the simplex $\Delta^{s-1} \subseteq \mathcal{S}$. Intersecting the image $M(\mathcal{H} \cap \Delta^{s-1}) \subseteq \mathcal{E}$ with the intersection polytope then yields the new modified intersection polytope. Algebraically, one just adds these relations as additional rows to the matrix defining the elemental composition map.

3.3. **Mixtures of Reactions.** In the chemical education literature, reactions admitting multiple balancings are often explained as a mixture of several other reactions. This is tantamount to writing a balance as a rational combination of balancings relative subpolytopes of the reaction and product polytopes.

3.3. *Example.* [14] Kolb explains the reaction:

$$3HClO_3 \rightarrow HClO_4 + Cl_2 + 2O_2 + H_2O$$

as a mixture of two reactions:

$$7HClO_3 \rightarrow 5HClO_4 + Cl_2 + H_2O$$
$$4HClO_3 \rightarrow 2Cl_2 + 5O_2 + 2H_2O$$

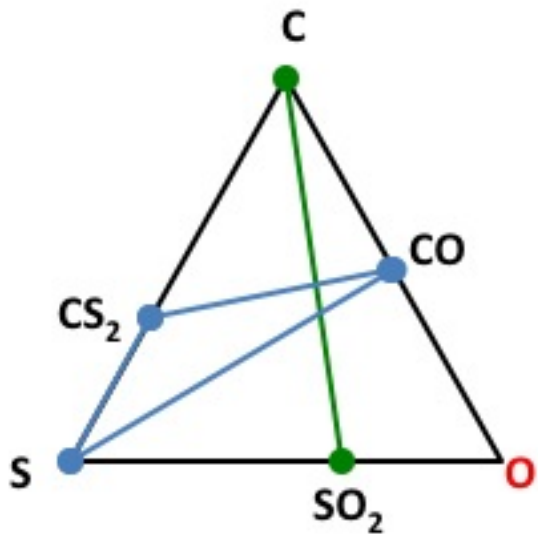

FIGURE 6. Reaction as Mixture of Reactions

where the first plus twice the second yields the desired overall reaction. From the polytopal diagram (see Figure 5), we see that $HClO_3$ is in the interior of the product polytope, and hence the reaction admits multiple balancings. The first reaction expresses the reactant via the subpolytope spanned by products $HClO_4, Cl_2, H_2O$ whereas the second via that spanned by products $Cl_2, O_2, H_2O$.

3.4. *Example.* [14] Kolb also explains the reaction:

$$3SO_2 + 7C \rightarrow CS_2 + S + 6CO$$

as a mixture of the reactions

$$SO_2 + 2C \rightarrow S + 2CO$$
$$2SO_2 + 5C \rightarrow CS_2 + 4CO$$

Again, from the polytopal diagram (see Figure 6), the existence multiple balancings is clear. Note that as in the preceding example, the particular choices for component reactions each admit unique balancings.

3.5. *Example.* [14] In the equation

$$NaClO_2 + NaClO \rightarrow NaClO_3 + NaCl$$

chlorine appears in four different oxidation states. Hence, application of the usual half-reaction method is somewhat confusing (see discussion in Subsection 4.3 below). Thankfully, the equation as written is already balanced, though it also admits multiple balancings. From the polytopal diagram (see Figure 7), we see that the species are actually collinear, since each has $Na, Cl$ in a $1 : 1$ ratio. Kolb explains this example as a combination of the reactions

$$3NaClO_2 \rightarrow 2NaClO_3 + NaCl$$
$$3NaClO \rightarrow NaClO_3 + 2NaCl$$

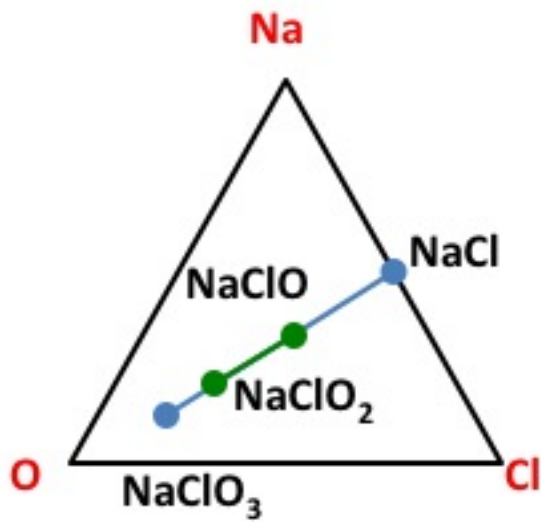

FIGURE 7. Reaction as Mixture of Reactions

In all the preceding examples, the choice of component reactions was chemically-motivated but also happened to admit unique balancings. Indeed, one may say they were chosen in part to be uniquely determined, since the paradigm is to explain multiple-balancings in terms of what are perceived to be "true" reactions, namely those admitting unique balancings. In fact, any reaction can be written as a rational combination of reactions with the reactants (resp. products) being subsets of the original reactants (resp. products) though as demonstrated in Example 3.4, we cannot require the subsets always to be proper. This follows from a simple observation: let conv() denote convex hull.

**3.6. Lemma.** *Suppose* $\mathrm{conv}(X) = \mathrm{conv}(R) \cap \mathrm{conv}(P)$ *for finite sets* $R, P$ *in some fixed Euclidean space. Then, there exist subsets* $R_i \subseteq R$ *and* $P_i \subseteq P$ *such that the intersections* $\mathrm{conv}(R_i) \cap \mathrm{conv}(P_i) = \{x_i\}$ *are singletons and any point* $x \in \mathrm{conv}(X)$ *is a convex combination of the points* $x_i$. *We cannot require* $R_i, P_i$ *to be simultaneously proper.*

*Proof.* We may assume that $X$ is the set of generating vertices for the intersection $\mathrm{conv}(R) \cap \mathrm{conv}(P)$. In general, for each $x_i \in X$, let $R_i, P_i$ be minimal subsets of $R, P$ respectively such that $x_i \in \mathrm{conv}(R_i) \cap \mathrm{conv}(P_i)$. If $x_i$ is an interior point of either $\mathrm{conv}(R), \mathrm{conv}(P)$ then a set of generating vertices for one of the two hulls may be needed. Hence, we cannot require $R_i, P_i$ to be necessarily proper subsets. If $x_i$ is a boundary point for both $\mathrm{conv}(R), \mathrm{conv}(P)$, then it is the intersection of a line segment of one hull and a face of the other hull. In either case the intersections $\mathrm{conv}(R_i) \cap \mathrm{conv}(P_i)$ are singletons. Since the $x_i$ generate $\mathrm{conv}(X)$, we are done. $\square$

Taking $R, P$ to be the sets of reactants and products, respectively and $\mathrm{conv}(X)$ to be the intersection polytope for the reaction, the we obtain the following:

**3.7. Proposition.** *Any chemical equation can be written as a rational combination of chemical equations admitting unique balance, with the reactants (resp. products)* *being subsets of the original reactants (resp. products) though we cannot require the subsets always to be proper.*

*Proof.* Refer to the proof of the preceding lemma. All that remains to show is that each $x_i$ encodes an unique balance. Note that any point of a finitely generated convex set is in the interior of a *simplex* generated by some (necessarily) minimal subset of the generating vertices. This simplex need not be a face of the polyhedron. If $x_i$ is an interior point, then the preceding observation shows that $x_i$ is in the interior of some simplex. If $x_i$ is a boundary point, then it is the intersection of a line segment and a face so that again by the opening observation, we may think of $x_i$ as the intersection of a line segment with a simplex. In either case, since the vertices of a simplex are linearly independent, by Equation (2.18), $x_i$ represents an unique balance.                                                                     □

This thus justifies the chemical paradigm of expressing chemical equations as mixtures of those admitting unique balance.

## 4. Balancing Equations Involving Charged Species

If charged species are involved, there are several ways we may procede: we may introduce (a) an additional component to index the charge or (b) fictitious "spectator ions" that allow us to balance the equation as if only neutral species are present. One important class of chemical reactions involving charged ions are *oxidation-reduction reactions*. Though these may be balanced by either of the two preceding methods, there is a commonly used "half-reaction method" that merits separate mathematical investigation since it is structurally different from the preceding methods.

4.1. **Component Indexing Charge.** We index charges as an additional component so that if $n$ elements are involved, then the species vectors are in $\mathbb{R}^{n+1}$ now. We use the convention that the charge is always indexed last and note that it may be negative. For example, the species $Ca^{2+}$ would have species vector $(1, 2)^T$ whereas $Cl^-$ would have $(1, -1)^T$. However, balancings still correspond to equalities of non-negative rational linear combinations, and *almost* everything above for both the algebraic and geometric methods remain valid. Evidently, the codomain of $R, P$ will be $\mathbb{R}^{n+1}$ so that the commutative diagram (2.17) should be altered accordingly, but the subtlety is that since we now allow negative entries in the last component, we no longer have a canonical choice for the slicing hyperplane.

If both lone $+/-$ charges (i.e. not bound with an element) are present, then no hyperplane always works because any hyperplane intersecting both the positive and negative $x_{n+1}$-rays contains the $x_{n+1}$-axis and hence passes through the origin. Of course, one can assume that the lone charges have been "canceled" so that only one kind remains. Alternatively, since in chemistry lone positive charges are denoted $H^+$, we may assume only lone negative charges are possibly present.

Recall, that the choice $\sum_{i=1}^{n} x_i = 1$ is canonical if charges do not appear by themselves without elements, e.g. free electrons $e^-$. When only one kind of free charge is involved (either positive or negative), we may be tempted to consider as a canonical hyperplane: $\pm x_{n+1} + \sum_{i=1}^{n} x_i = 1$, where the sign depends on the parity of the free charge. Unfortunately, as the following example illustrates, this hyperplane will *not* necessarily intersect both the "reactant" and "product" cones:

4.1. *Example.* Consider the reaction:

$$HNO_2 \rightarrow NO_2 + H^+ + e^-$$

Because a lone negative charge is a specie, we use the hyperplane $-x_4 + \sum_{i=1}^{3} x_i = 1$ (remember that charge is indexed last). However, the ray corresponding to $H^+$ is of the form $(k, 0, 0, k)$ for real $k > 0$ and thus does not intersect the hyperplane.

Even without a canonical slicing hyperplane, there still exist (many!) hyperplanes which slice both "reactant" and "product" cones of a given reaction. If we insist on working with polytopes instead of cones, we need only slice with any such hyperplane, albeit non-canonical.

4.2. **Spectator Ions.** One method chemists utilize for balancing equations involving charged species is to introduce fictitious *spectator ions* (e.g. $Q^+, X^-$) which are formal symbols bound with the charged species so that the equation no longer involves charged species. After a balance is found, the spectator ions are deleted. The following example illustrates this approach:

4.2. *Example.* Consider the *unbalanced* equation:

$$Zn + NO_3^- + H^+ \rightarrow Zn^{2+} + NH_4^+ + H_2O \qquad \text{(unbalanced)}$$

Using spectator ions $Q^+, X^-$ we instead balance

$$Zn + NO_3Q + HX \rightarrow ZnX_2 + NH_4X + H_2O + QX$$

Note the additional inclusion of $QX$ among the products so that $Q$ will be present on both sides. This yields an unique balance:

$$4Zn + NO_3Q + 10HX \rightarrow 4ZnX_2 + NH_4X + 3H_2O + QX$$

Hence, the desired balance to our original equation is

$$4Zn + NO_3^- + 10H^+ \rightarrow 4Zn^{2+} + NH_4^+ + 3H_2O$$

The geometric explanation is simple: Let $\mathcal{E}$ be the non-negative span of the elements involved in the reaction. Then the image of the species simplex lies in $\mathcal{E} \times \mathbb{R}$ where the last coordinate indexes charge. Thinking of $\mathbb{R}$ as the union of the non-negative and non-positive rays, we have $\mathcal{E} \times \mathbb{R} = \mathcal{E} \times (\mathbb{R}_{\geq 0} \cup \mathbb{R}_{\leq 0})$ as sets (simply by twisting one of the rays orthogonally). Identifying the latter two components as the non-negative span of $X, Q$ respectively, we see that the codomain may be taken as the simplex spanned by the elements together with the spectator ions. Each specie is represented by a point in either the $\mathcal{E}Q, \mathcal{E}X$-hyperplanes or the point $QX$ in the $QX$-plane, though the reactant and product cones are not restricted to these hyperplanes. In fact, the introduction of $QX$ is necessary in situations where the reactant or product polytope lie in different boundary faces of the orthorant of $\mathcal{E} \times \mathbb{R}_{\geq 0} \times \mathbb{R}_{\leq 0}$ or only one does while the other is in the interior. See Figure 8.

4.3. **Half-Reaction Method.** One important class of chemical reactions are called *oxidation-reduction* reactions. They are characterized by changes in the *oxidation states* of elements and often involve $H^+, OH^-, H_2O$ as species. In fact, these species are sometimes not explicitly specified until an equation is balanced. For chemical details, see any general chemistry textbook (e.g. Olmsted[19]). One common approach to balancing oxidation-reduction reactions is the "half-reaction" or "ion-electron" method. This method first divides the reactant and product species into two "half-reactions" corresponding to an "oxidation" and a "reduction" reaction. This selection is determined by chemical considerations. The half-reactions are separately balanced using $H^+, OH^-, H_2O, e^-$ and then linearly combined so as to cancel unbound elections $e^-$. The chemical considerations assure that unbound electrons in the two half-reactions will occur on opposite sides of the equation and hence are amenable to an unique cancellation via linear combination. Thus, the result of the half-reaction method is unique iff each half-reaction admits an unique balance.

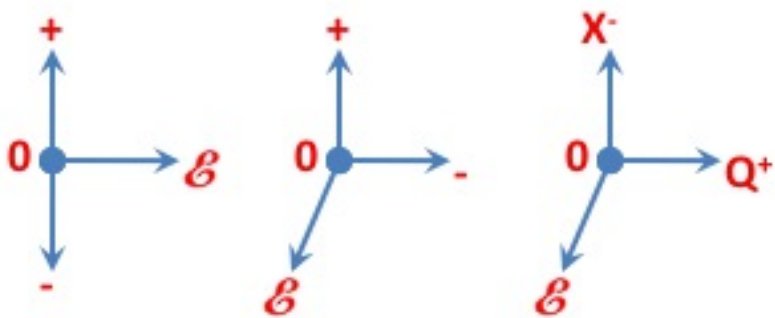

FIGURE 8. Geometric Explanation for Spectator Ions

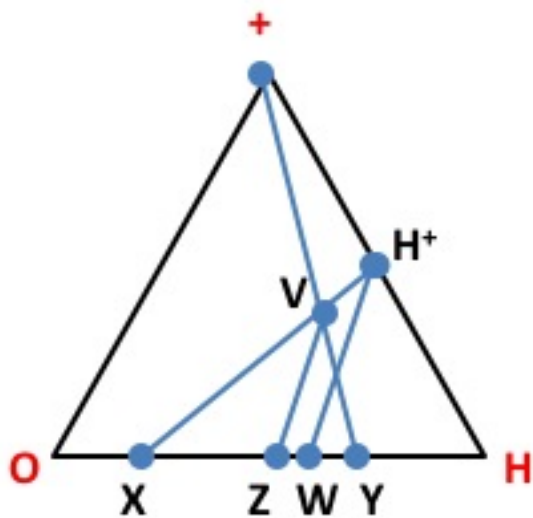

FIGURE 9. Geometry of Balancing Redox Half-Reaction: Steps (3) and (4).

Each half-reaction is balanced using the following recipe:

(1) balance all elements except $H, O$, and the charge
(2) balance oxygen $O$ with $H_2O$
(3) balance hydrogen $H$ with $H^+$ ions

(4) (for basic solutions): cancel $H^+$ using the equality $H_2O = H^+ + OH^-$

(5) balance charge with electrons $e^-$

The geometry of steps (3) and (4) are illustrated in Figure 9. Symbols $X, Y$ represent the complexes (i.e. each individual side of the equation) after step (2), shown by their projections in directions O,H only. The diagram shows the case where complex $X$ is deficient in $H$ and hence balanced with $H^+$ resulting in $V$. Symbol $W$ denotes water $H_2O$ and $Z$ is the result of (4) (recall both charge directions are projected out). Line segments $\overline{H^+W}$ and $\overline{VZ}$ are parallel. The remaining step (5) balances the two lifts of $Z$ (as reactant and product complexes) from the face spanned by the elements into its join with the charge (i.e. spectator ion) directions.

Notice that the above recipe provides an unique balance to a half-reaction iff the projection onto the orthogonal complement of the subspace indexing $H, O$, and charge admits an unique balancing. All steps subsequent to the first are uniquely determined and provide a sequence of lifts of the partial solution into the subspaces indexed by $O, H$, and charge, respectively.

The following example illustrates the "half-reaction method":

**4.3.** *Example.* [19, Ex 17.7, p827] Consider the following reaction in *basic solution*:

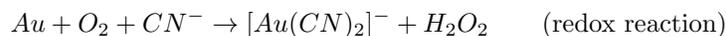

$$Au + O_2 + CN^- \rightarrow [Au(CN)_2]^- + H_2O_2 \qquad \text{(redox reaction)}$$

By chemical principles one identifies that gold $Au$ is oxidized by cyanide ions $CN^-$

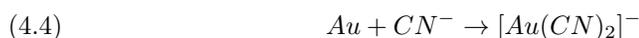

$$(4.4) \qquad\qquad Au + CN^- \rightarrow [Au(CN)_2]^-$$

and oxygen gas $O_2$ is reduced to hydrogen peroxide $H_2O_2$:

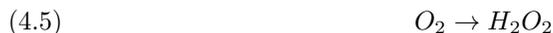

$$(4.5) \qquad\qquad O_2 \rightarrow H_2O_2$$

The oxidation half-reaction (Equation 4.4) is readily balanced as

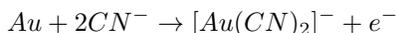

$$Au + 2CN^- \rightarrow [Au(CN)_2]^- + e^-$$

and the reduction half-reaction (Equation 4.5) as

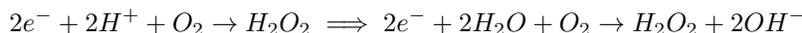

$$2e^- + 2H^+ + O_2 \rightarrow H_2O_2 \implies 2e^- + 2H_2O + O_2 \rightarrow H_2O_2 + 2OH^-$$

since the reaction is taking place in basic solution. Hence, the overall reaction is

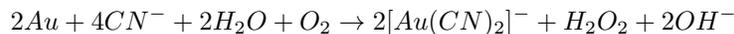

$$2Au + 4CN^- + 2H_2O + O_2 \rightarrow 2[Au(CN)_2]^- + H_2O_2 + 2OH^-$$

However, the directness of the recipe belies a subtlety: the intersection cone associated to the *unbalanced* reduction half-reaction

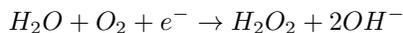

$$H_2O + O_2 + e^- \rightarrow H_2O_2 + 2OH^-$$

has dimension 2, so that both this half-reaction and the total reaction admit multiple balancings. The subtlety lies in the fact that for basic solutions the recipe does not balance using $H_2O, OH^-$ directly as in the preceding equation, but instead uses $H^+$ and then converts via $H_2O = H^+ + OH^-$. For the preceding example, this choice constrains the balanced reduction reaction to have $H_2O, OH^-$ in a $1:1$ ratio. In effect, it slices the intersection cone for the half-reaction with the hyperplane defined by equal coefficients for $H_2O, OH^-$. Indeed, the one-parameter family of balancings for the reduction action yields a one-parameter family of balancings for overall reaction:

$$(4.6) \quad (1-t)Au + 2(1-t)CN^- + \left(\frac{1+t}{2}\right)H_2O + \left(\frac{1+t}{4}\right)O_2 \to$$
$$(1-t)[Au(CN)_2]^- + tH_2O_2 + (1-t)OH^-$$

for $t \in [0,1]$. The output of the "half-reaction" method corresponds to $t = 1/3$. In particular, the reaction:

$$(4.7) \quad 2Au + 4CN^- + 4H_2O + 2O_2 \to 2[Au(CN)_2]^- + 3H_2O_2 + 2OH^-$$

from using the reduction half-reaction ($t = 3/5$):

$$2e^- + 4H_2O + 2O_2 \to 3H_2O_2 + 2OH^-$$

is not obtainable via our two chosen half-reactions since $H_2O, OH^-$ are not in a $1:1$ ratio.

One may wonder if Equation 4.7 were obtainable using the "half-reaction" recipe relative a different set of half-reactions. This is not the case: for our redox reaction, we may classify half-reactions by whether $[Au(CN)_2]^-$ is a product. If so, then $Au, CN^-$ need be present as reactants to admit a balance. The only other possible product combination is $H_2O_2$ alone; since $Au, CN^-$ cannot be included among the reactants without requiring $[Au(CN)_2]^-$ among the products, we have only four possible half-reactions:

| Unbalanced | Balanced |
|---|---|
| $Au + CN^- \to [Au(CN)_2]^-$ | $Au + 2CN^- \to [Au(CN)_2]^- + e^-$ |
| $Au + CN^- + O_2 \to [Au(CN)_2]^-$ | $3e^- + 2H_2O + Au + 2CN^- \to [Au(CN)_2]^- + 4OH^-$ |
| $Au + CN^- \to [Au(CN)_2]^- + H_2O_2$ | $Au + 2CN^- + 2OH^- \to [Au(CN)_2]^- + H_2O_2 + 3e^-$ |
| $O_2 \to H_2O_2$ | $2e^- + 2H_2O + O_2 \to H_2O_2 + 2OH^-$ |

where each has been balanced using the recipe for balancing half-reactions. One checks that no pair of half-reactions yields Equation 4.7. Hence, *the half-reaction method cannot necessarily find all possible balances.*

Recall that the half-reaction method outputs a balancing iff there exist two half-reactions for which step (1) above is realizable, i.e. the projection onto the orthogonal complement of the subspace indexing $H, O$, and charge admits a balancing. This orthogonal complement is equivalently the span of all non $H, O$-elements. Suppose a balancing for the overall reaction exists; then the projection onto the aforesaid complement is non-empty. Now, denote the set of "reactant" and "product" species by $R, P$ respectively. Hence, step (1) is realizable iff there exists decompositions $R = R_1 \cup R_2$ and $P = P_1 \cup P_2$ such that $\text{conv}(R_i) \cap \text{conv}(P_i) \neq \emptyset$ for $i = 1, 2$. Here, conv() denotes the convex hull of sets. The following example shows that we cannot require the subsets to be disjoint:

**4.8.** *Example.* [14] Consider the following *unbalanced* oxidation-reduction reaction:

$$P_2I_4 + P_4 + H_2O \to PH_4I + H_3PO_4 \qquad \text{(unbalanced)}$$

If we order $P$ before $I$ in indexing elements, under the projection onto the span of $P, I$ the coordinates for the reactants are $(2,4)^T, (4,0)^T$ and that for products $(1,1)^T, (1,0)^T$. A glance at the polytope diagram (see Figure 10) shows that we cannot require both $R_1, R_2$ and $P_1, P_2$ to be disjoint pairs.

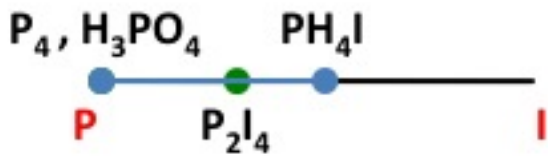

Figure 10. Cannot Partition Reactants and Products into Disjoint Pairs

The following lemma assures us that otherwise existence follows readily:

**4.9. Lemma.** *Let* $\mathrm{conv}(X) = \mathrm{conv}(R) \cap \mathrm{conv}(P)$ *where* $\mathrm{conv}(X)$ *has dimension* $c \leq d$ *for non-empty finite sets* $R, P, X \subseteq \mathbb{R}^d$ *where we may assume each set individually to be convex linearly independent.*

*Let* $x \in \mathrm{conv}(X)$. *Then, there exist decompositions* $R = R_1 \cup R_2$ *and* $P = P_1 \cup P_2$ *such that defining* $H_i = \mathrm{conv}(R_i) \cap \mathrm{conv}(P_i)$ *for* $i = 1, 2$ *we have* $x \in \mathrm{conv}(H_1 \cup H_2)$. *Furthermore, if* $c$ *is positive then we may require* $H_i$ *to be non-empty.*

*Proof.* We first provide an intuitive argument. Since $\mathrm{conv}(X)$ is convex, excepting the trivial case when $\mathrm{conv}(X)$ is a singleton, we may see $x$ by peering through any boundary facet of $\mathrm{conv}(X)$. By shifting ourselves slightly, we can place $x$ in the interior of a background facet of $\mathrm{conv}(X)$. Then, $x$ is in the convex hull of these two facets of $\mathrm{conv}(X)$.

To formalize the above argument, we place $\mathrm{conv}(X)$ in a $c$-ball (where $c$ is the dimension of $\mathrm{conv}(X)$) such that the center of the ball is in the interior of $\mathrm{conv}(X)$. Taking the radial projection outwards from the center point yields a cellular decomposition of the boundary sphere. Choose any cell. By appropriate choice of point $p$ within this cell, the line through $p, x$ meets the sphere at another point $q$ interior to some other cell on the sphere (here we use the fact that $\mathrm{conv}(X)$ is positive-dimensional). These two cells identify two facets $H_1, H_2$ of $\mathrm{conv}(X)$ such that their convex hull includes $x$. Since each face of $\mathrm{conv}(X)$ is the intersection of some faces of $\mathrm{conv}(R), \mathrm{conv}(P)$ there exist subsets $R_i, P_i$ such that $H_i = \mathrm{conv}(R_i) \cap \mathrm{conv}(P_i)$. Finally, note that since $H_i$ are facets of $\mathrm{conv}(X)$, at least one of $R_i, P_i$ is a proper subset for each $i = 1, 2$. $\square$

**4.10.** *Remark.* In Example 4.3, the projection onto the complement of $H, O$ and charge yields only the species $Au, CN, Au(CN)_2$. The projected reactant and product polytopes meet at a singleton. Hence, we are not assured two non-trivial half-reactions in the space spanned by $Au, C, N$.

**4.11. Proposition.** *If an oxidation-reduction reaction admits a balance, then there exist decompositions of the reactants and products, each into two (not necessarily disjoint) groups so that application of the half-reaction method yields a balance relative the given reactants and products. By appropriate choice of decompositions, this balance can be made to reflect the elemental proportions of all non-O,H elements in a given initial balance.*

*Proof.* As mentioned previously, we need only show that we can decompose the reaction into half-reactions for which step (1) of the recipe is realizable. Since the reaction admits a balance, the intersection cone remains non-empty after projecting out the $H, O$ and charge-directions. Hence, the projections of the reactant and product polytopes satisfy the hypotheses of the lemma. We let $R_i, P_i$ be as assured by the lemma. Then, $H_i = R_i \cap P_i$ is non-empty so that step (1) is realizable for each half-reaction. □

The half-reaction method has been criticized by some chemists because sometimes the choice of oxidation/reduction reactions is unclear and even chemically fictitious[15]. As discussed above, the method has mathematical drawbacks in that the recipe may obscure the presence of multiple balancings and may not capture all possible balances. However, at least the preceding proposition assures us that if a balancing exists, then the "half-reaction method" can generate a valid balance, though the half-reactions may be chemically fictitious.

## 5. Reaction Mechanisms and Stoichiometric Restrictions

In this section we discuss the relationship between reaction mechanisms and stoichiometric restrictions. Given a chemical reaction, a reaction mechanism is a sequence of steps that together elucidate how the reactants of the overall reaction are converted into the products of the overall reaction. Each step is called an *elementary chemical reaction* and represents the most fundamental molecular transformations.

### 5.1. From Mechanism to Linear Stoichiometric Relations.
A given reaction mechanism has implications for observed stoichiometries of the involved species as illustrated in the following example:

5.1. *Example.* [22] Let $S_1, S_2, S_3$ be species and consider the following reaction mechanism:

$$S_1 \rightarrow S_2 + S_3$$
$$S_1 + S_2 \rightarrow 2S_3$$

Letting $[S_i]$ denote (time-dependent) concentrations, the theory of chemical kinetics implies that

$$\frac{d[S_1]}{dt} = -X_1 - X_2$$
$$\frac{d[S_2]}{dt} = X_1 - X_2$$
$$\frac{d[S_3]}{dt} = X_1 + 2X_2$$

where the $X_i$ depend only on the $i$-th elementary reaction in the mechanism. Eliminating the $X_i$ yields the single relation:

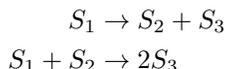

$$\frac{d}{dt}(3[S_1] + [S_2] + 2[S_3]) = 0$$

which constrains the observed stoichiometries.

We may formalize this as follows: Order the species involved and let $N$ be a matrix with columns indexed by the steps of the mechanism and the rows indexed by species. We will follow the convention in the literature of negative entries for reactant species and positive entries for product species. For the preceding example, we would have

$$N = \begin{pmatrix} -1 & -1 \\ 1 & -1 \\ 1 & 2 \end{pmatrix}$$

Then, elimination of the $X_i$ is tantamount to finding elements of $NS(N^T)$ which we may think of as the space of stoichiometric restrictions. Each vector in this subspace defines a linear relation on the concentration of species that integrates to a stoichiometric restriction which in the chemical literature is called a *mass-conservation equation*. (The terminology arises from the assumption of constant volume during the course of the reaction.) In particular, mass-conservation equations are defined with respect to a given *set* of reactions. Note that though $d[S_i]/dt = X_i$, since we do not assume a particular form of $X_i$ we need not assume mass-action kinetics. This shows that for given initial concentrations $[S_i]$, the resulting flow is constrained to lie in an affine space linear isomorphic to $\mathcal{N} = \mathrm{CS}(N)$, the orthogonal complement to $NS(N^T)$. Further information can be obtained if we assume mass-action kinetics, whence we obtain an ODE system of polynomials. The literature on these systems is vast (see for example either Horn and Jackson[13] or Érdi and Tóth[9]).

In addition, chemists also define *element conservation equations*. These are defined for any given *single* reaction. Given the elemental proportion matrix $M$ for a reaction, since the rows index elements, we may require their conservation in the course of a reaction. Thus, the set of all element-conservation equations corresponds to the row space $\mathrm{RS}(M)$.

Up to now our discussion has not assumed any elemental composition for the species nor need there be one. The existence of such implies a (non-negative) elemental proportion matrix $M$ such that $MN = 0$ because each reaction indexed by $N$ is balanced. In particular, we may assume the rows of $M$ index only elements present so that the row space $\mathrm{RS}(M)$ must contain a positive vector. Since $\mathrm{NS}(N^T) \supseteq RS(M)$, a necessary condition for the existence of $M$ is that $\mathrm{NS}(N^T)$ contains a positive vector. In the chemical literature, such matrices $N$ are known as *conservative* [18]. The sufficiency of this condition is readily checked. Furthermore, given an overall reaction and a reaction mechanism for it, since the rows of $M$ index elements, the relation $\mathrm{RS}(M) \subseteq \mathrm{NS}(N^T)$ shows that any elemental-conservation equation is a mass-conservation equation but not vice-versa (see also Smith and Missen[18]). Letting $\mathcal{X}, \mathcal{S}, \mathcal{E}$ denote the spaces spanned by elementary reactions, species, and elements respectively and thinking of $N, M$ as linear maps, we have

$$\mathcal{X} \xrightarrow{N} \mathcal{S} \xrightarrow{M} \mathcal{E}$$

and the discrepancy between mass- and element-conservation equations is described by the homology $(\ker M)/(\operatorname{im} N)$ at $\mathcal{S}$.

In a reaction, there are usually short-lived species called *intermediates* that are difficult to observe. Under a steady-state hypothesis, or if we consider only completed reactions, or just assuming that they cannot be detected, the space of *observed* linear dependencies is instead the orthogonal projection of $\mathrm{NS}(N^T)$ onto the span of observable species. We denote this space $\pi_{\mathcal{K}} \, \mathrm{NS}(N^T)$.

5.2. **Further Conclusions from a Given Mechanism.** In this subsection we show how several chemically interesting problems lead to the problem of counting lattice points in polytopes, a much-studied and active research area in combinatorics.

5.2.1. *Reactions Consistent with Mechanism.* Let $N$ be an integer matrix associated with a collection of (elementary) reactions, and write $\mathcal{N} = \mathrm{CS}_{\geq 0}(N)$ for the *cone* determined by its non-negative span. Let $K$ be the set of known (or observable) species and $\mathcal{K}$ their span. For a chemical equation $\mathbf{c} \in \mathcal{K}$ we will say that $\mathbf{c}$ is *consistent with $N$* or that $N$ *is a mechanism for $\mathbf{c}$* provided $\mathbf{c} \in \mathcal{N}$.

For a given $N$, the set of consistent chemical equations is thus the set of lattice points in $\mathcal{N} \cap \mathcal{K}$ modulo dilation, or alternatively the points of the projective subset $\mathbb{P}_{\mathbb{Q}}(\mathcal{N} \cap \mathcal{K})$. Suppose the cone $\mathcal{N} \cap \mathcal{K}$ is non-trivial. Then, $\mathbb{P}_{\mathbb{Q}}(\mathcal{N} \cap \mathcal{K})$ is either a singleton representing an integer lattice point or contains infinitely-many such representatives. One can justify the latter claim by noting the infinitude of rational vectors in any non-trivial projective neighborhood of an integral vector.

Alternatively, via Ehrhart theory the number of lattice points in integral dilations of a rational $d$-polytope $P$ are counted by a quasi-polynomial $E_{\overline{P}}$ called the *Ehrhart quasi-polynomial* (see [7],[8]). That is, there exists some $N > 0$ and degree $d$-polynomials $f_0, \ldots, f_{N-1}$ such that $E_{\overline{P}}(t) = f_i(t)$ for $t \equiv i \mod N$. Lattice points in the relative interior are counted by another quasi-polynomial $E_P(t)$ where the two quasi-polynomials are related by $E_{\overline{P}}(t) = (-1)^d E_P(-t)$ where $d = \dim P$. Hence, if $d > 1$, then $E_P$ grows at least quadratically so that $\mathbb{P}_{\mathbb{Q}}(\mathcal{N} \cap \mathcal{K})$ contains infinitely-many points. Chemically, this implies that there is either an unique chemical equation consistent with a given mechanism or there are infinitely many. However, since we do not expect to observe overall reactions built from many instances of each elementary step, in practice we see but finitely-many consistent equations.

Suppose we want to consider overall reactions containing at most $t$ species (counting multiplicities). Then, we can count the lattice points within the polytope formed by bounding the cone $\mathcal{N} \cap \mathcal{K}$ with the cross-polytopes $C_t^d = \{\mathbf{z} \in \mathbb{R}^d \mid \sum |z_i| = t\}$. This is computationally feasible since there exist polynomial-time algorithms to compute the number of lattice points in polytopes. See deLoera[7] or Barvinok[2],[3] and the programs LattE[6] and barvinok[24]. (The output of the latter is technically a piece-wise step-polynomial, a generalization of quasi-polynomials.) However, we should keep in mind that the counts from lattice point enumeration in $\mathcal{S}$ may include multiples of a given balance.

In light of the above discussion, we may be tempted to define a partial order on the interior points of the cone $\mathcal{N} \cap \mathcal{K}$ via the total number of known species involved (counting multiplicity), namely $\sum_K s_i$. However, there exist lattice points on arbitrarily large cross-polytopes such that their convex hull with the origin does not include any interior lattice point. For example, taking $i, j, k$ unique indices, let $\mathbf{x} = t\mathbf{e}_i - \mathbf{e}_j$ and $\mathbf{y} = t\mathbf{e}_i - \mathbf{e}_k$ for positive integer $t$. The convex hull $\mathrm{conv}(\{\mathbf{0}, \mathbf{x}, \mathbf{y}\})$ contains no interior point. Thus, relative the $l^1$-norm there may not be a minimum consistent reaction.

5.2. *Remark.* We may also want to consider consistent reactions with mechanisms involving no more than a fixed number of reactions. We then would have to work in reaction space $\mathcal{X}$ by counting lattice points in the intersection of cross-polytopes and $\mathcal{X}_{\geq 0} \cap N^{-1}(\mathcal{N} \cap \mathcal{K})$ where $\mathcal{X}_{\geq 0}$ is the non-negative orthant in $\mathcal{X}$.

5.3. *Remark.* This observation on lattice points of cross-polytopes also applies to studying the polyhedron $\mathcal{Q} \subseteq \mathcal{S}$ that parametrizes balances of a given chemical equation. Hence, using the sum of the coefficients in a balance may not be sufficient to single out a unique "simplest" balance. However, we may use the Ehrhart theory to estimate the minimal total number of species involved in the overall reaction (counting multiplicity) by intersecting $\mathcal{Q}$ with cross-polytopes $C_t^s$ in $\mathcal{S}$.

5.4. *Remark.* We express Craciun and Pantea's result on *confoundable* reaction networks[5] using our framework. Two reaction networks are confoundable if they yield the same kinetic differential equations under the assumption of mass-action kinetics.

Let $N$ be the equations of a reaction network expressed as a matrix (columns indexing reactions, rows species), $\mathbf{S}$ the vector of specie concentrations, and $k_i$ the forward rate-constant for the $i$-th reaction. Let $\mathbf{x}$ be a vector of dimension equal to the number of elementary reactions in the mechanism, and suppose that the entries of $\mathbf{x}$ depend on a function of the reactant (i.e. source) complexes alone. Then, $d\mathbf{S}/dt = N \operatorname{diag}(k_i)\mathbf{x}$. We may assume the components of $\mathbf{x}$ and the columns of $N$ are sorted by reactant complex so that $N \operatorname{diag}(k_i)$ has only as many columns as distinct source complexes and the dimension of $\mathbf{x}$ is the number of distinct reactant complexes. (Just add together columns of $N \operatorname{diag}(k_i)$ indexed by reactions having identical reactant complex so each column of $N \operatorname{diag}(k_i)$ is a cone on $N$ parametrized by the $k_i$ of reactions with that reactant complex.) Thus, a mechanism associates to each reactant complex a cone. Then, two networks yield the same kinetic differntial equations provided they have identical set of reactant complexes and for a fixed reactant complex, the cone so indexed for each of the two mechanisms intersect non-trivially. This is Craciun and Pantea's result, though without assuming mass-action kinetics.

5.5. *Example* (Decomposition of Azomethane). We consider the decomposition of azomethane:

$$(5.6) \qquad 5C_2H_6N_2 \rightarrow 3N_2 + CH_4 + C_2H_6 + C_3H_8N_2 + C_4H_{12}N_2$$

A reaction mechansim is given by Missen and Smith[18] which involves intermediates that we will label $X, Y, Z$. Under the ordering

$$C_2H_6N_2, N_2, CH_4, C_2H_6, C_3H_8N_2, C_4H_{12}N_2, X, Y, Z$$

the mechanism has associated matrix

$$(5.7) \qquad N = \begin{pmatrix} -1 & -1 & 0 & 0 & -1 & 0 \\ 1 & 0 & 0 & 0 & 0 & 0 \\ 0 & 1 & 0 & 0 & 0 & 0 \\ 0 & 0 & 1 & 0 & 0 & 0 \\ 0 & 0 & 0 & 1 & 0 & 0 \\ 0 & 0 & 0 & 0 & 0 & 1 \\ 2 & -1 & -2 & -1 & -1 & -1 \\ 0 & 1 & 0 & -1 & 0 & 0 \\ 0 & 0 & 0 & 0 & 1 & -1 \end{pmatrix}$$

To count consistent overall reactions where no more than a fixed number of species are present in the equation, we consider the intersections $(\operatorname{cone}(\mathbf{0}, N) \cap \mathcal{K}) \cap C_t^9$ where $C_t^9$ are cross-polytopes. For example, the intersection with $C_6^9$ is a 6-polytope in $\mathcal{S} = \mathbb{R}^9$ having 14 vertices. By BARVINOK we find that this polytope has 35 lattice points so that there are 35 consistent overall reactions where there are no more than 6 species (counting multiplicity).

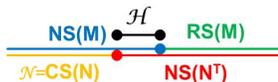

FIGURE 11. Relations among subspaces of $\mathcal{S}$

5.2.2. *Algebraic Representations of a Reaction.* Let $\mathbf{c} \in \mathcal{K}$ be a given chemical equation with mechanism $N$. We will call any non-negative integral solution to $N\mathbf{x} = \mathbf{c}$ an *algebraic representation* for $\mathbf{c}$. Algebraic representations do not necessarily correspond to chemical pathways within a given mechanism because the presence of intermediates in elementary reactions means that the order of reactions in a mechanism is important. For example let $I_1, I_2$ be intermediates, and suppose the following pair of equations:

$$S_1 + I_1 = S_2 + I_2 \qquad S_3 + I_2 = S_4 + I_1$$

comes from some mechanism. Then their sum is an algebraic representation for $S_1 + S_3 = S_2 + S_4$, but the two reactions alone cannot constitute a chemically viable mechanism unless one of the intermediates is present at the start of the reaction. Further discussion of this restriction is postponed until Subsection 5.4. For now we note that the number of algebraic representations is given by the number of non-negative lattice points in the polyhedron determined by $N\mathbf{x} = \mathbf{c}$. As mentioned above, this is computationally feasible to determine. This number is finite iff $\mathrm{NS}(N)$ has no non-negative element, or equivalently that $\mathbf{0}$ is not in the convex hull of the columns of $N$. If infinite, we may still study representations involving no more than $t$-steps in total by slicing the non-negative solution cone of $N\mathbf{x} = \mathbf{c}$ in $\mathcal{X}$ with the hyperplane $\sum_{i=1}^{x} x_i = t$ where $x = \dim \mathcal{X}$ the number of columns of $N$.

The case of unique algebraic representation corresponds to polytopes $\{\mathbf{x}|N\mathbf{x} = \mathbf{c}\}$ with unique interior lattice point. Because we are working over $\mathbb{Z}$, this may happen even if $\ker N \neq 0$. For example, the polytope corresponding to the intersection of the hyperplane $\{(x,y)|x + y = 2\}$ with the non-negative quadrant is a line segment with an unique interior lattice point. Polytopes in $\mathbb{R}^2$ with unique interior lattice point have been classified up to unimodular equivalence in Rabinowitz[20], but seems unknown in higher dimensions. Already, there exist lattice simplices in $\mathbb{R}^d$ with unique interior lattice point but having more than $2^{2^{d-1}}$ boundary lattice points (see Zaks, Perles, and Wills[25]).

5.8. *Example.* Again, consider the mechanism for the decomposition of azomethane given in Example 5.5. The convex hull of the columns of $N$ determines a convex 5-polytope in $\mathcal{S} = \mathbb{R}^9$. Since this polytope does not contain $\mathbf{0}$, there are only finitely-many non-negative solutions to $N\mathbf{x} = \mathbf{c}$ for any $\mathbf{c} \in \mathcal{S}$. Using $\mathbf{c} = (-5, 3, 1, 1, 1, 1)^T$ from the reaction (5.6), we find that the polytope parametrizing non-negative solutions is actually just a point, namely $(3, 1, 1, 1, 1, 1)^T$. Thus, this is the unique non-negative integral combination of elementary reactions of the mechanism $N$ yielding the overall reaction (5.6).

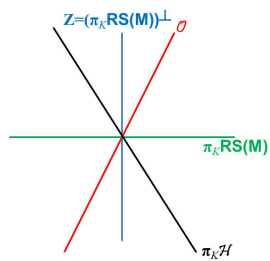

FIGURE 12. Relation among subspaces of $\mathcal{K}$. The possibility of non-trivial intersection between $\mathcal{O}$ and $Z = (\pi_{\mathcal{K}} \operatorname{RS}(M))^{\perp}$ has been suppressed in the Diagram.

5.3. **Mechanisms Consistent with Observed Relations.** We now consider an inverse problem. Suppose we know all the species (including intermediates) involved in a reaction with unknown mechanism; denote their span by $\mathcal{S}$. A set of linear dependencies are observed which span a space $\mathcal{O}$ (obtained via multivariate linear regression of data, for example). What can we say about mechanisms $N$ compatible with these restrictions and in particular of the image space $\mathcal{N} = \operatorname{im} N$? (Note here $\mathcal{N}$ represents a linear space contra the cone in the preceding subsection.) We know that for a given chemical equation $\mathbf{c}$, the moduli of mechanisms for $\mathbf{c}$ is given by the set of all integer matrices $N$ such that $\mathbf{c} \in \mathcal{N}$.

Figures 11 and 12 summarize the relations among the subspace now to be defined. Let $K, U$ be the sets of known and unknown (intermediates) species, respectively and $S = K \cup U$. Write $\mathcal{K}, \mathcal{U}, \mathcal{S}$ for the spaces they span and equip $\mathcal{S}$ with the canonical inner product. Then $\mathcal{S} = \mathcal{K} \perp \mathcal{U}$. Now, write $\ker M = \mathcal{N} \perp \mathcal{H}$ where we think of $\mathcal{H}$ as the homology at $\xrightarrow{N} \mathcal{S} \xrightarrow{M}$. Then, $\mathcal{S} = \mathcal{N} \perp \mathcal{H} \perp \operatorname{RS}(M)$ so that

$$\mathcal{H} \perp \operatorname{RS}(M) = \operatorname{NS}(N^T) = \pi_{\mathcal{K}} \operatorname{NS}(N^T) \perp \pi_{\mathcal{U}} \operatorname{NS}(N^T)$$

where $\pi_{\mathcal{K}}, \pi_{\mathcal{U}}$ denote orthogonal projection onto $\mathcal{K}, \mathcal{U}$ respectively. Write $\mathcal{O}$ for the space of observed dependencies. We will not assume that this constitutes all possible linear dependencies, hence we have

(5.9) $$\pi_{\mathcal{K}} \mathcal{H} \perp \pi_{\mathcal{K}} \operatorname{RS}(M) = \pi_{\mathcal{K}} \operatorname{NS}(N^T) = \mathcal{O} \perp W$$

for some subspace $W$. In particular, we see that

(5.10) $$\dim \pi_{\mathcal{K}} \operatorname{NS}(N^T) \geq \dim(\mathcal{O} + \pi_{\mathcal{K}} \operatorname{RS}(M))$$

where the right-hand side may be computed *without knowledge of any intermediates*. If we specify intermediates, we can do much better. From Equation (5.9) we

have

$$(5.11) \qquad [\mathcal{O} + W] \subseteq [\pi_\mathcal{K}\mathcal{H}] \subseteq \frac{\mathcal{K}}{\pi_\mathcal{K}\,\mathrm{RS}(M)}$$

and since $\mathcal{K} = \pi_\mathcal{K}\,\mathrm{NS}(M) + \pi_\mathcal{K}\,\mathrm{RS}(M)$ we have

$$(5.12) \qquad \frac{\pi_\mathcal{K}\,\mathrm{NS}(M)}{\pi_\mathcal{K}\,\mathrm{NS}(M) \cap \pi_\mathcal{K}\,\mathrm{RS}(M)} \cong \frac{\mathcal{K}}{\pi_\mathcal{K}\,\mathrm{RS}(M)} \cong (\pi_\mathcal{K}\,\mathrm{RS}(M))^\perp =: Z \subset \pi_\mathcal{K}\,\mathrm{NS}(M)$$

so that we must choose a subspace $X$ such that

$$(5.13) \qquad \mathrm{proj}_Z \mathcal{O} \subseteq X =: \mathrm{proj}_Z \mathcal{H}$$

where we choose a proper superset when we believe that $W \neq 0$, i.e. that $\mathcal{O}$ do not represent all possibly observable stoichiometric restrictions. Regardless, we have

$$(5.14) \qquad \pi_\mathcal{K}\mathcal{H} = X + (\pi_\mathcal{K}\,\mathrm{NS}(M) \cap \pi_\mathcal{K}\,\mathrm{RS}(M))$$
$$\supseteq \mathrm{proj}_Z \mathcal{O} + (\pi_\mathcal{K}\,\mathrm{NS}(M) \cap \pi_\mathcal{K}\,\mathrm{RS}(M))$$

Since $\mathcal{H} \subseteq \mathrm{NS}(M)$, to lift $\pi_\mathcal{K}\mathcal{H}$ back up to $\mathrm{NS}(M)$ we need to study the kernel of the restriction of $\pi_\mathcal{K}$ to $\mathrm{NS}(M)$. We denote this map $\varphi = \pi_\mathcal{K}\colon \mathrm{NS}(M) \to \mathcal{K}$. Writing $M = (M_K \quad M_U)$ where $M_K, M_U$ are the submatrices of $M$ indexed by known and intermediate species, respectively, we see that $\mathrm{NS}(M)$ is the centralizer of the maps $Mi_\mathcal{K}\pi_\mathcal{K}, -Mi_\mathcal{U}\pi_\mathcal{U}$ so that $\ker \varphi = i_\mathcal{U}\,\mathrm{NS}(M_U)$. Thus,

$$(5.15) \qquad \mathcal{H} = \pi^{-1}X + \pi^{-1}(\pi_\mathcal{K}\,\mathrm{NS}(M) \cap \pi_\mathcal{K}\,\mathrm{RS}(M)) + i_\mathcal{U}\,\mathrm{NS}(M_U)$$
$$\supseteq \pi^{-1}\,\mathrm{proj}_Z \mathcal{O} + \pi^{-1}(\pi_\mathcal{K}\,\mathrm{NS}(M) \cap \pi_\mathcal{K}\,\mathrm{RS}(M)) + i_\mathcal{U}\,\mathrm{NS}(M_U)$$

In the case $\mathrm{NS}(N^T) = \mathcal{O}$, we have $X = \mathrm{proj}_Z \mathcal{O}$ so that the possible choices for $\mathcal{H}$ (and hence $\mathcal{N}$ since $\mathcal{N} = (\mathcal{H} \oplus \mathrm{RS}(M))^\perp$) are parametrized by subspaces of

$$(\pi_\mathcal{K}\,\mathrm{NS}(M) \cap \pi_\mathcal{K}\,\mathrm{RS}(M)) \perp i_\mathcal{U}\,\mathrm{NS}(M_U)$$

with dimension bounded by $\dim(\mathcal{O}/Z)$. We illustrate these ideas with two examples:

5.16. *Example.* Again, consider the decomposition of azomethane given by Equation (5.6) of Example 5.5 above. With the species ordered as in that example and ordering the elements $C, H, N$ in that order, the associated elemental proportion matrix is

$$M = \begin{pmatrix} 2 & 0 & 1 & 2 & 3 & 4 & 1 & 2 & 3 \\ 6 & 0 & 4 & 6 & 8 & 12 & 3 & 5 & 9 \\ 2 & 2 & 0 & 0 & 2 & 2 & 0 & 2 & 2 \end{pmatrix}$$

where the intermediates $X, Y, Z$ have molecular formulae $CH_3, C_2H_5N_2, C_3H_9N_2$ respectively. Suppose the space $\mathcal{O} \subset \mathcal{K} = \mathbb{R}^6$ of observed stoichiometric restrictions has a basis

$$(-1, -3, 1, 2, 0, 1)^T \qquad (2, 4, 0, -2, 0, 0)^T \qquad (0, 0, -1, 0, 1, 0)^T$$

Let $Z = (\pi_\mathcal{K}RS(M))^{\perp\mathcal{K}}$. We compute that $\dim(\mathcal{O}/Z) = 0$. Thus, assuming $\mathcal{O}$ includes all possible stoichiometric restrictions, we may set $\mathcal{H} = 0$.

Also, we find that $\dim \pi_\mathcal{K}RS(M) \cap \pi_\mathcal{K}NS(M) = 3$ and that $\mathrm{NS}(M_U) = 0$ so that the possible $\mathcal{N}$ are parametrized by

$$\mathcal{N} = (\mathrm{span}(\mathbf{s}) + \mathrm{RS}(M))^\perp \qquad \mathbf{s} \in \pi^{-1}(\pi_\mathcal{K}RS(M) \cap \pi_\mathcal{K}NS(M))$$

5.17. *Example* (Oxidation of Formaldehyde). Missen and Smith provide a mechanism in [18] for the oxidation of formaldehyde. If we order the elements by $H, C, O, Co, +$ and the species by

$$CH_2O, Co^{3+}, Co^{2+}, H^+, CO, H_2O, CH_2O_2, H_2, CH_2OCo^{3+}, CHO, OH, H$$

where the last four species are intermediates, we have the associated elemental proportion matrix

$$(5.18) \qquad M = \begin{pmatrix} 2 & 0 & 0 & 1 & 0 & 2 & 2 & 2 & 2 & 1 & 1 & 1 \\ 1 & 0 & 0 & 0 & 1 & 0 & 1 & 0 & 1 & 1 & 0 & 0 \\ 1 & 0 & 0 & 0 & 1 & 1 & 2 & 0 & 1 & 1 & 1 & 0 \\ 0 & 1 & 1 & 0 & 0 & 0 & 0 & 0 & 1 & 0 & 0 & 0 \\ 0 & 3 & 2 & 1 & 0 & 0 & 0 & 0 & 3 & 0 & 0 & 0 \end{pmatrix}$$

The given mechanism has associated matrix

$$(5.19) \qquad N = \begin{pmatrix} -1 & 0 & 0 & 1 & 0 & -1 & -1 & -1 \\ -1 & 0 & -1 & 0 & 0 & 0 & 0 & 0 \\ 0 & 1 & 1 & 0 & 0 & 0 & 0 & 0 \\ 0 & 1 & 1 & 0 & 0 & 0 & 0 & 0 \\ 0 & 0 & 1 & 0 & 0 & 0 & 0 & 0 \\ 0 & 0 & 0 & -1 & -1 & 1 & 0 & 0 \\ 0 & 0 & 0 & 0 & 1 & 0 & 1 & 0 \\ 0 & 0 & 0 & 0 & 0 & 0 & 0 & 1 \\ 1 & -1 & 0 & 0 & 0 & 0 & 0 & 0 \\ 0 & 1 & -1 & -1 & -1 & 1 & 0 & 1 \\ 0 & 0 & 0 & 1 & 0 & -1 & -1 & 0 \\ 0 & 0 & 0 & 0 & 1 & 0 & 1 & -1 \end{pmatrix}$$

so that the space $\mathcal{O} = \pi_{\mathcal{K}} \operatorname{NS}(N^T) \subset \mathcal{K} = \mathbb{R}^8$ has a basis

$$(1,0,0,0,1,0,0,1)^T \qquad (0,0,0,1,-2,0,0,0)^T$$
$$(0,1,0,0,2,0,0,0)^T \qquad (0,0,0,0,0,1,0,1)^T$$
$$(0,0,1,0,-2,0,0,0)^T \qquad (0,0,0,0,0,0,1,-1)^T$$

Suppose we started knowing only the species and the above space $\mathcal{O}$. Set $Z = (\pi_{\mathcal{K}} \operatorname{RS}(M))^{\perp_{\mathcal{K}}}$ and let $\operatorname{proj}_Z$ denote the orthogonal projection onto $Z$. We find that $\dim(\mathcal{O}/Z) = 1$ where

$$\operatorname{proj}_Z \mathcal{O} = \operatorname{span}\left(1, \frac{-12}{23}, \frac{12}{23}, \frac{12}{23}, \frac{-49}{23}, \frac{-26}{23}, \frac{26}{23}, \frac{-29}{23}\right)^T$$

Assuming all stoichiometric restrictions are included in $\mathcal{O}$, i.e. $\mathcal{O} = \pi_{\mathcal{K}} \operatorname{NS}(N^T)$, we have $\pi_{\mathcal{K}} \mathcal{H} = \operatorname{proj}_Z \mathcal{O}$. A basis for $\pi_{\mathcal{K}} \operatorname{NS}(M) \cap \pi_{\mathcal{K}} \operatorname{RS}(M)$ is given by

$$(2,-1,0,0,0,0,0,2)^T \qquad (1,0,0,0,1,0,1,0)^T$$
$$(-1,0,0,0,-1,1,0,0)^T \qquad (0,2,1,1,0,0,0,0)^T$$

and this represents the uncertainty in determining $\pi_{\mathcal{K}} \mathcal{H}$. One checks that $\pi_{\mathcal{K}}$ is injective on $\operatorname{NS}(M)$ so that $\operatorname{proj}_Z \mathcal{O} + (\pi_{\mathcal{K}} \operatorname{NS}(M) \cap \pi_{\mathcal{K}} \operatorname{RS}(M))$ lifts uniquely to $\operatorname{NS}(M)$. Hence, compatible subspaces $\mathcal{N} = (\mathcal{H} \perp \operatorname{RS}(M))^{\perp}$ are parametrized by the 4-dimensional space $\pi_{\mathcal{K}} \operatorname{NS}(M) \cap \pi_{\mathcal{K}} \operatorname{RS}(M)$. Setting $\pi_{\mathcal{K}} \mathcal{H} = \operatorname{proj}_Z \mathcal{O}$ yields

$$\mathcal{H} = \operatorname{span}\left(\frac{40}{23}, \frac{26}{23}, 1, 1, \frac{-29}{23}, \frac{-14}{23}, \frac{26}{23}, \frac{-26}{23}, \frac{-49}{23}, \frac{12}{23}, \frac{-12}{23}, 1\right)^T$$

However, using the mechanism $N$ given above by Missen and Smith yields

$$\mathcal{H} = \operatorname{span}\left(1, \frac{-47}{64}, \frac{15}{32}, \frac{15}{32}, \frac{-75}{32}, \frac{-5}{128}, \frac{7}{4}, \frac{-101}{128}, \frac{17}{64}, \frac{-43}{64}, \frac{-219}{128}, \frac{315}{128}\right)^T$$

The difference of their projections into $\mathcal{K}$ is

$$\left(\frac{761}{12928}, \frac{-1573}{6464}, \frac{-9}{404}, \frac{-9}{404}, \frac{-4379}{12928}, \frac{13249}{12928}, \frac{4435}{6464}, \frac{1285}{3232}\right)^T$$

which one verifies is in the subspace $\pi_{\mathcal{K}} \operatorname{NS}(M) \cap \pi_{\mathcal{K}} \operatorname{RS}(M)$.

The dimensions of important subspaces in Examples 5.16 and 5.17 are summarized in the table below:

| Space | Ex 5.17 | Ex 5.16 |
|---|---|---|
| $\operatorname{NS}(M)$ | 7 | 6 |
| $\operatorname{RS}(M)$ | 5 | 3 |
| $\mathcal{K}$ | 8 | 6 |
| $\pi_{\mathcal{K}} \operatorname{NS}(M)$ | 7 | 6 |
| $\pi_{\mathcal{K}} \operatorname{RS}(M)$ | 5 | 3 |
| $\pi_{\mathcal{K}} \operatorname{NS}(M) \cap \pi_{\mathcal{K}} \operatorname{RS}(M)$ | 4 | 3 |
| $\mathcal{O}$ | 6 | 3 |
| $\mathcal{O}/\pi_{\mathcal{K}} \operatorname{RS}(M)$ | 1 | 0 |
| $\pi_{\mathcal{U}} \operatorname{RS}(M)$ | 5 | 3 |

5.4. **From Subspaces $\mathcal{N}$ to Mechanisms.** Two tasks remain in obtaining reaction mechanisms consistent with observed stoichiometric restrictions. The first is convex-geometric: we seek spanning sets in $\mathcal{N}$ whose positive hull contains the overall reaction. This ensures algebraic representability. The second is order-theoretic: these algebraic representations must then be subject to the constraint on the appearance of intermediates discussed in Subsection 5.2.2.

The first task seems computationally expensive: Elementary reactions have usually no more than two (or at most three) molecules of a given specie so that we can restrict attention to those lattice points in $\mathcal{N} \cap [-3, 3]^s$ where $s = \dim \mathcal{S}$ having both positive and negative entries. However, this grows exponentially with the number of species involved. Perhaps some savings may result from working projectively and this may also be helpful enumerating spanning sets, but on this the author is uncertain. Nevertheless, as explained in Subsection 5.2.2, algebraic representatives may be obtained from the lattice points.

The second task is more tractable. Let $X \subset \mathcal{X}$ be a finite set of reactions, $u$ the number of distinct intermediates among the reactions in $X$, and $\Delta^u$ the $u$-dimensional simplex (as an abstract simplicial complex). We may assume there are no redundacies in $X$. We label the vertices $K, U_1, \ldots, U_u$. Define *set* maps $R, P \colon X \to \Delta^u$ mapping a reaction to the simplex indexed by its reactants and products respectively, with knowns collapsed to $K$. For example, if we have knowns $K_1, K_2$ and intermediates $U_1, U_2$, the maps $R, P$ take the reaction $K_1 + U_1 \to K_2 + U_2$ to the 1-simplices $[K, U_1]$ and $[K, U_2]$ respectively.

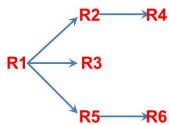

Figure 13. Precedence relations among the reactions in the mechanism for the decomposition of azomethane. See Example 5.20 and subsequent discussion.

Let $V(\cdot)$ denote the set of vertices of a simplicial complex, and for each set $A$ let $\Delta(A)$ denote the simplex (as an abstract simplicial complex) with vertices indexed by the elements of $A$. For any *non-empty* subcomplex $Y \subseteq \Delta^u$ define

$$\varphi(Y) := (\Delta \circ V)(X \cup PR^{-1}(Y))$$

The coface to the vertex $K$ in the simplex $\varphi(Y)$ indexes all intermediates involved (either as reactants or products) in the reactions indexed by $X$. We set $\varphi(\{\emptyset\}) = \{K\}$. Since we assume $X$ is finite, the iterates

$$(\varphi \circ \cdots \circ \varphi)(\{\emptyset\})$$

eventually stabilize. In particular the $U$-indices of this complex index the intermediates that can possibly arise through reactions in set $X$. Note also that if a reaction $x \in X$ occurs then there is some sequence of reactions $x_1, \ldots, x_N$ where the reactants of $x_1$ are known species and the intermediate reactants of $x_j$ are among the intermediate products of $x_i$ for $i < j$. Thus, $x \in (R^{-1} \circ \varphi^{N+1})(\{\emptyset\})$ so that if a reaction occurs then it is an element of $(R^{-1} \circ \varphi^i)(\{\emptyset\})$ for some $i > 0$. The converse is readily checked. Finally, note that the sets $(R^{-1} \circ \varphi^i)(\{\emptyset\})$ for sucessive $i \geq 1$ define a partial order on the elements in $X$ the set of reactions. In particular, reactions not involved in this partial order cannot occur.

5.20. *Example.* Consider again the decomposition of azomethane from Example 5.5 where we denoted the intermediates $X, Y, Z$. Since $\mathcal{N} = \mathrm{NS}(M)$ we may compute that there are 116 non-zero intergal vectors in the intersection $\mathrm{NS}(M) \cap [-1, 1]^9$, or equivalently 58 distinct lines. It seems computationally prohibitive to enumerate spanning sets for $\mathcal{N}$ among these without further heuristics.

Nonetheless, choosing as a particular example the set of equations given by the mechanism $N$ in Equation (5.7), we have

| Rxn | $P(\cdot)$ | $R(\cdot)$ |
|:---:|:---:|:---:|
| 1 | K | [K,X] |
| 2 | [K,X] | [K,Y] |
| 3 | X | K |
| 4 | [X,Y] | K |
| 5 | [K,M] | Z |
| 6 | [X,Z] | K |

We compute that $\varphi(K) = \Delta(\{K, X\})$ and that $\varphi^2(K) = \Delta^3$. The associated poset is given in Figure 13.

Thus, we see that an algebraic representation is order realizable iff it is equal to $(R^{-1} \circ \varphi^i)(\{\emptyset\})$ for some $i$.

**5.21.** *Remark.* We have only considered stoichiometric properties of reactions. In particular we have not incorporated stereochemical and thermodynamic considerations and such may help reduce the search space in determining spanning sets for computing algebraic representatives. Furthermore, much of the discussion requires knowledge of the intermediate species which would need to be guessed by chemical intuition. Regardless, it is hoped that the above procedure provides an attempt at the computational discovery of reaction mechanisms.

## 6. Appendix: Mathematical Justifications for Geometric Viewpoint

In this section we justify several mathematical assertions from Section 2. First, here is the justification of the linear algebraic fact invoked in the discussion relation the polyhedron $\mathcal{Q}$ and the intersection cone $\mathcal{I}$:

Let $S = \mathrm{CS}(A_{n \times a})$ and $T = \mathrm{CS}(B_{n \times b})$ and consider the $n \times (a+b)$ matrix $M$ with columns from $A, B$ respectively. Then, $S + T = \mathrm{CS}(M)$.

**6.1. Proposition.** $\mathrm{nullity}(M) - \dim(S \cap T) = \mathrm{nullity}(A) + \mathrm{nullity}(B)$.

*Proof.* By the Rank-Nullity Theorem for $M$:

$$\dim(\mathrm{CS}(M)) + \dim(\mathrm{NS}(M)) = a + b$$

By choosing a basis for $S \cap T$, separately extending to bases for $S, T$ and then invoking inclusion-exclusion:

$$\dim(S + T) + \dim(S \cap T) = \dim(S) + \dim(T)$$

Subtracting these two equations and using the Rank-Nullity Theorem for $A, B$ yields

$$\mathrm{nullity}(M) - \dim(S \cap T) = \mathrm{nullity}(A) + \mathrm{nullity}(B)$$

$\square$

Next, we provide geometric justifications for the statements from Section 2 that for a given choice of "reactants" and "products":

(1) A non-empty intersection polytope necessarily contains a rational point.
(2) Each rational point in the intersection polytope corresponds to some balancing of the chemical equation.

Algebraic arguments essentially follow from "chasing" Diagram 2.17.

The former follows readily from the following two results:

**6.2. Proposition.** *The intersection of two polytopes with rational vertices is a polytope with rational vertices.*

*Proof.* Since the polytopes have rational vertices, each supporting hyperplane may be chosen to be of the form $\sum n_i x_i = h$ for rational $n_i, h$. Now, each vertex in the intersection polytope is given as the unique intersection locus of certain supporting hyperplanes, hence the unique solution of a system $N\mathbf{x} = \mathbf{h}$ where $N, \mathbf{h}$ have rational entries. Thus, each vertex is rational. $\square$

6.3. **Proposition.** *A positive-dimensional polytope with rational vertices contains a rational point in its relative interior.*

*Proof.* Let $d$ be the dimension of the polytope and $\{\mathbf{x}_i\}$ any $d+1$ generating vertices whose affine hull is $d$-dimensional. Any convex linear combination $\sum_{i=1}^{d} t_i \mathbf{x}_i$ with $t_i$ rational and $0 < t_i < 1$ yields a point with the desired properties. $\qquad\square$

6.4. **Corollary.** *A non-empty intersection polytope necessarily contains a rational point. If the intersection polytope is positive-dimensional, we can find a rational point in its relative interior.*

*Proof.* The "reactant" and "product" polytopes have rational vertices. Hence, the intersection polytope has rational vertices, and by the preceding proposition, we are done. $\qquad\square$

6.5. **Proposition.** *Each rational point in the intersection polytope corresponds to some balancing of the chemical equation.*

*Proof.* It suffices to show that a rational point in the convex hull of rational points is expressible as a rational linear combination of those points. We proceed by induction on the number of generating vertices and may assume without loss of generality that the generators are a minimal set for the given hull. The base case of a line segment determined by its two endpoints is evident. Suppose the statement is true for points in the convex hull of at most $d$ rational points. Let $\mathbf{p}$ be a rational point in the convex hull of $d + 1$ rational points. We may assume that the convex hull is not generated by any proper subset of the $d + 1$ points. Let $\mathbf{q}$ be one of the generating points. By using a rational frame based at $\mathbf{q}$, we may assume the ambient space is $\mathbb{R}^d$. The $d$ generating points besides $\mathbf{q}$ determine a $(d-1)$-hyperplane $H$ of the form $\mathbf{n} \cdot \mathbf{x} = \mathbf{b}$ where $\mathbf{n}, \mathbf{b}$ are rational and $\mathbf{n}$ is normal to $H$. Then, $\mathbf{n} \cdot \mathbf{p}$ and $\mathbf{n} \cdot \mathbf{q}$ are rational numbers so that by taking dot product of $\mathbf{n}$ with the orthogonal decompositions

$$(6.6) \qquad \mathbf{p} = \mathbf{s} + p\mathbf{n} \qquad \mathbf{q} = \mathbf{t} + q\mathbf{n}$$

where $\mathbf{s}, \mathbf{t}$ are in hyperplane $H$, we conclude that $p, q \neq 0$ are rational. Finally,

$$(6.7) \qquad \mathbf{p} = \mathbf{s} + \frac{p}{q}(\mathbf{q} - \mathbf{t}) = \left(\mathbf{s} - \frac{p}{q}\mathbf{t}\right) + \frac{p}{q}\mathbf{q}$$

a rational linear combination as desired. $\qquad\square$

*E-mail address*: `chuang.jerchin@gmail.com`